\documentclass[aps,pra,showpacs,floatfix,superscriptaddress,twocolumn,bibnotes]{revtex4-1}
\usepackage{graphicx}
\usepackage{amsmath}
\usepackage{amssymb}
\usepackage{bm}
\usepackage{color}
\usepackage{dcolumn}
\usepackage{multirow}
\usepackage[caption=false]{subfig}
\usepackage{hyperref}

\begin{document}

\title{Modeling motional energy spectra and lattice light shifts in optical lattice clocks}

\newcommand{\NIST}{
National Institute of Standards and Technology, Boulder, Colorado, USA
}

\newcommand{\CU}{
Department of Physics, University of Colorado, Boulder, Colorado, USA
}

\newcommand{\GTRI}{
Georgia Tech Research Institute, Atlanta, GA, USA
}

\author{K. Beloy}
\email{kyle.beloy@nist.gov}
\affiliation{\NIST}

\author{W. F. McGrew}
\affiliation{\NIST}
\affiliation{\CU}

\author{X. Zhang}
\affiliation{\NIST}

\author{D. Nicolodi}
\affiliation{\NIST}

\author{R. J. Fasano}
\affiliation{\NIST}
\affiliation{\CU}

\author{Y. S. Hassan}
\affiliation{\NIST}
\affiliation{\CU}

\author{R. C. Brown}
\altaffiliation[present address: ]{\GTRI}
\affiliation{\NIST}

\author{A. D. Ludlow}
\affiliation{\NIST}
\affiliation{\CU}

\date{\today}

\begin{abstract}
We develop a model to describe the motional (i.e., external degree of freedom) energy spectra of atoms trapped in a one-dimensional optical lattice, taking into account both axial and radial confinement relative to the lattice axis. Our model respects the coupling between axial and radial degrees of freedom, as well as other anharmonicities inherent in the confining potential. We further demonstrate how our model can be used to characterize lattice light shifts in optical lattice clocks, including shifts due to higher multipolar (magnetic dipole and electric quadrupole) and higher order (hyperpolarizability) coupling to the lattice field. We compare results for our model with results from other lattice light shift models in the literature under similar conditions.
\end{abstract}


\maketitle

\newcommand{\later}{\ensuremath{\spadesuit}}

\section{Introduction}

Since their inception less than two decades ago~\cite{Kat02,KatTakPal03}, optical lattice clocks have been demonstrated for a number of atomic species~\cite{TakKat03,BarHoyOat06,YiMejMcF11,KulFimZip15,GolFedTre19,YamSafGib19} in several laboratories across the world, as well as in compact, transportable form~\cite{KolGroVog17,GroKolVog18}. State-of-the-art performance has steadily improved, with key metrics of accuracy, stability, and reproducibility now being realized at or below the $10^{-18}$ fractional level~\cite{UshTakDas15,McGZhaFas18,BotKedOel19}. The very premise of optical lattice clocks implies a strong perturbation to the atoms, as interaction with the lattice light serves as the mechanism for atomic confinement. By operating the lattice at or near the ``magic'' frequency, the confining potential is largely independent of the clock state. As a consequence, spectroscopy on the clock transition can be performed largely free of effects of atomic motion. Additionally, the atomic confinement permits long interrogation times~\cite{SchBroMcG17,CamHutMar17} and the implementation of well-controlled environments~\cite{BelHinPhi14,UshTakDas15,BelZhaMcG18}, which will be especially important for the continued progress of these clocks.

Lattice light shifts in optical lattice clocks can be interpreted as residual effects of atomic motion stemming from small differences in the confining potentials of the clock states. A difference in potentials can arise due to a deviation from the magic frequency or from higher order/multipolar~\cite{TaiYudOvs08} coupling to the lattice field. While the importance of lattice light shifts was certainly appreciated from the beginning, the characterization of these shifts has remained an active field of study throughout the evolution of optical lattice clocks~\cite{BruLeTBai06,BarStaLem08,KatHasIli09,WesLodLor11,OvsPalTai13,KatOvsMar15,OvsMarPal16,BroPhiBel17,UshTakKat18,NemJorYan19}.

In this paper, we develop a model to describe the motional energy spectra of atoms trapped in a one-dimensional optical lattice. We further demonstrate how this model can be extended to characterize lattice light shifts in optical lattice clocks. While this work is motivated by our interest in optical lattice clocks, it could prove beneficial for other applications employing optical lattice-trapped atoms~\cite{BloZol06}.

\section{The optical lattice potential}

To start, we restrict our attention to the dominant second order electric dipole ($E1$) coupling of the atoms to the optical lattice field, encapsulated by the $E1$ polarizability $\alpha_{E1}$. In terms of cylindrical coordinates $(\rho,\varphi,z)$, we take the optical lattice potential to be
\begin{equation}
U(\rho,z)=
-\left(\frac{\mathcal{E}_0}{2}\right)^2\alpha_{E1}e^{-\kappa^2\rho^2}\cos^2\left(kz\right),
\label{Eq:potential}
\end{equation}
where $\mathcal{E}_0$ is the peak electric field amplitude in the lattice. Here $\kappa=\sqrt{2}/w$ and $k=2\pi/\lambda$, where $w$ is the $1/e^2$ power radius of the lattice and $\lambda$ is the lattice wavelength. $\alpha_{E1}$ implicitly depends on $\lambda$ and is assumed to be positive. (Elsewhere in this work, we refer to lattice frequency in favor of lattice wavelength, with the understanding that they are simply related by the speed of light.) In practice, tunneling between lattice sites can be suppressed by working with sufficiently deep potentials and sufficiently cold atomic samples, as well as aligning the lattice along gravity~\cite{LemWol05}. We exclude tunneling in our analysis as described below. Lastly, we assume $\kappa/k\ll1$ and $D\gtrsim10E_R$, where $D=\left(\mathcal{E}_0/2\right)^2\alpha_{E1}$ is the potential depth and $E_R=\hbar^2k^2/2m$ is the recoil energy associated with absorption of a lattice photon. In the expression for recoil energy, $\hbar$ is the reduced Planck constant and $m$ is the atomic mass. As a quantitative example, the Yb optical lattice clocks described in Ref.~\cite{McGZhaFas18} operate with $\kappa/k\approx10^{-3}$ and $D\approx50E_R$. As we are specifically interested in lattice trapped atoms in this work, we only concern ourselves with states having energy less than zero.

\section{The harmonic oscillator potential and the perturbative treatment}
\label{Sec:HOapprox}

In the regime $\kappa\rho,kz\ll 1$, the potential can be approximated by a series expansion truncated at terms second order in the coordinates, 
$U(\rho,z)\approx U^\mathrm{HO}(\rho,z)$, where
\begin{equation}
U^\mathrm{HO}(\rho,z)= D\left(-1+\kappa^2\rho^2+k^2z^2\right).
\label{Eq:HOpot}
\end{equation}
As this merely represents decoupled radial and axial harmonic oscillator potentials (along with an energy offset), the corresponding wave functions and energies are known. Specifically, the energies read
\begin{equation}
\begin{aligned}
E^\mathrm{HO}_{n_\rho l n_z}=&
-D+2\sqrt{DE_R}(\kappa/k)\left(2n_\rho+\left|l\right|+1\right)
\\&
+2\sqrt{DE_R}\left(n_z+\frac{1}{2}\right),
\end{aligned}
\label{Eq:HOenergies}
\end{equation}
where $n_\rho$ and $n_z$ take on non-negative integer values and $l$ takes on integer values. The quantum number $l$ specifies the $z$-component of angular momentum in units of $\hbar$. The corresponding wave functions are provided in Appendix \ref{Sec:HOwf}.

Given our assumptions, namely $\kappa/k\ll 1$, we see that the spectrum $E^\mathrm{HO}_{n_\rho l n_z}$ is dense in both $n_\rho$ and $l$. Consequently, we can express the spectrum as a density of states for a given $n_z$. From Eq.~(\ref{Eq:HOenergies}), we deduce this to be (see Appendix \ref{Sec:HODOS})
\begin{equation}
G^\mathrm{HO}_{n_z}(E)=\frac{\left(\kappa/k\right)^{-2}}{4DE_R}
\left[E+D-2\sqrt{DE_R}\left(n_z+\frac{1}{2}\right)\right],
\label{Eq:HODOS}
\end{equation}
where the argument $E$ denotes energy. Only non-negative values are physical, with $G^\mathrm{HO}_{n_z}(E)$ understood to be zero when the right-hand-side of this expression returns negative values (corresponding to an absence of states below a certain energy).

For the deepest bound states, the harmonic oscillator energies and wave functions may give an entirely adequate description of the energies and wave functions of the full potential. Moving higher into the spectrum, the wave functions become less localized, and it may be necessary to consider higher order terms in the series expansion of the potential. These terms include radial and axial anharmonic corrections, as well as cross-dimensional terms coupling the radial and axial degrees of freedom. If the effects of these additional terms are sufficiently small, then they may be treated as perturbations to the harmonic oscillator potential. In this approach, one must choose which terms in the series expansion of the potential are to be included through which orders of perturbation theory. Given that there are simple analytical expressions for the matrix elements of $\rho^2$ and $z^2$ between the harmonic oscillator states (see Appendix~\ref{Sec:HOwf}), as well as a simple analytical expression for the harmonic oscillator energies (Eq.~(\ref{Eq:HOenergies})), analytical expressions for the energy corrections can, in principle, be obtained for arbitrary choices. However, while the resulting expressions are simple in the sense that they only involve elementary mathematical operations, they quickly become cumbersome with increasing number of expansion terms and perturbation orders. Thus, the perturbative approach is only practical if convergence with respect to the series expansion and perturbation order is demonstrated to be sufficiently rapid.

\begin{figure*}[tb]
\subfloat{\includegraphics[width=160pt]{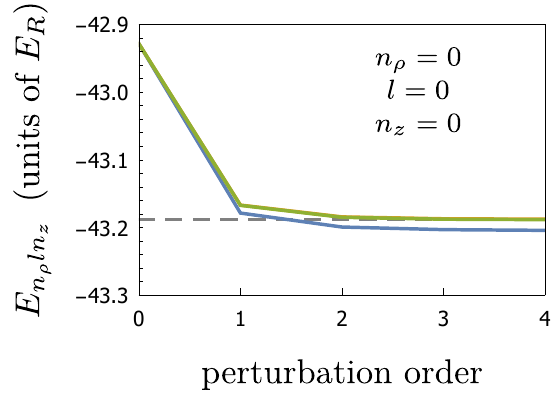}}%
\hspace{15pt}%
\subfloat{\includegraphics[width=160pt]{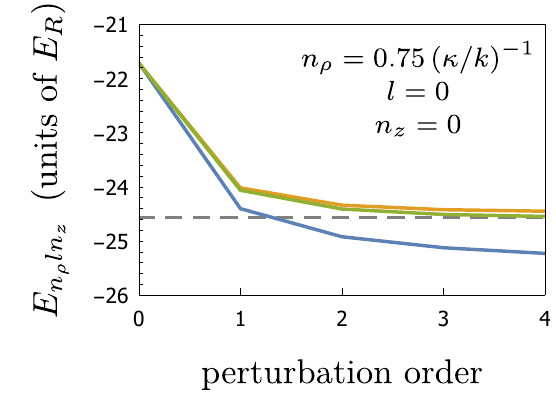}}%
\hspace{15pt}%
\subfloat{\includegraphics[width=160pt]{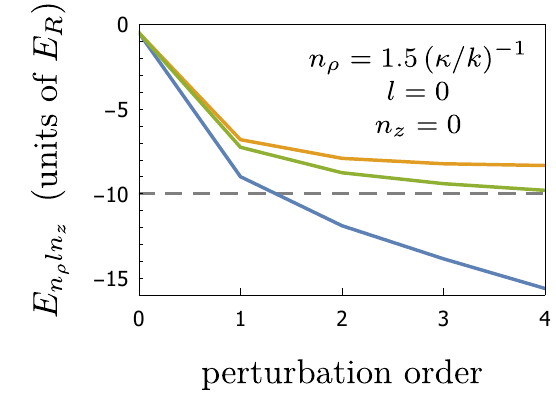}}%
\\
\subfloat{\includegraphics[width=160pt]{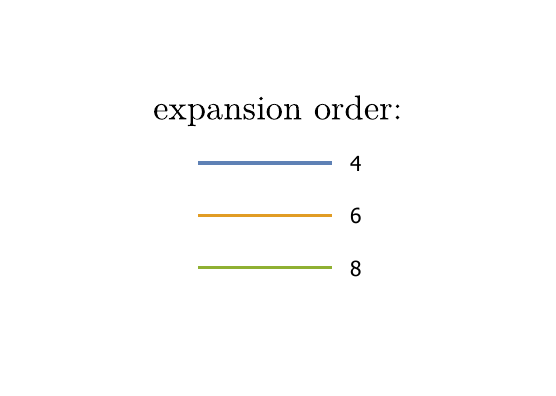}}%
\hspace{15pt}%
\subfloat{\includegraphics[width=160pt]{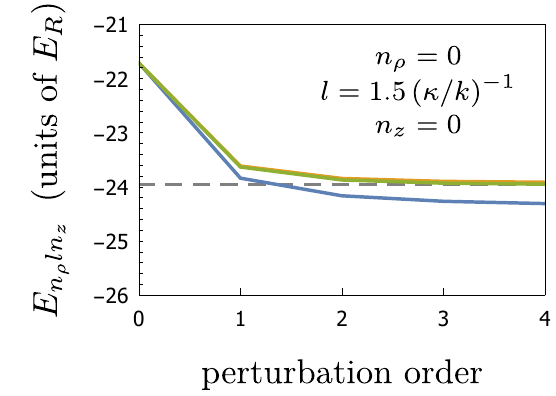}}%
\hspace{15pt}%
\subfloat{\includegraphics[width=160pt]{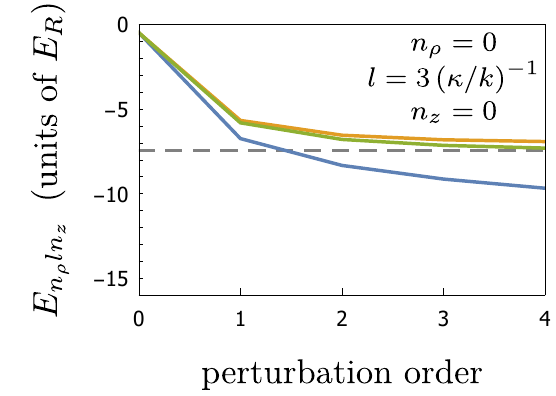}}%
\caption{Motional energies for $D=50E_R$ and $\kappa/k\rightarrow0$ determined using the perturbative approach (Section~\ref{Sec:HOapprox}) and the BO+WKB model (Sections~\ref{Sec:BO} and \ref{Sec:WKB}). Considered are the ground state (left panel), two states from the middle region of the spectrum (center panels), and two states from the upper region of the spectrum (right panels). The states have definite values of $\left(\kappa/k\right)n_\rho$ and $\left(\kappa/k\right)l$, with $n_z=0$ in all cases. For the perturbative approach, full lines depict the progression with respect to perturbation order, where the perturbation is taken to include all terms in the series expansion of $U\left(\rho,z\right)-U^\mathrm{HO}\left(\rho,z\right)$ through fourth (blue), sixth (yellow), and eighth (green) order in the coordinates. Fourth order in the coordinates, for example, refers to terms $\propto\rho^pz^q$ with $p+q=4$. The horizontal dashed line in each panel marks the energy according to the BO+WKB model.}
\label{Fig:HOPTWKB}
\end{figure*}

To explore this quantitatively, we compute energy corrections for select motional states assuming a depth $D=50E_R$ and the limit $\kappa/k\rightarrow0$. We consider the ground state ($E^\mathrm{HO}_{n_\rho l n_z}=-42.9E_R$), as well as states from the middle and upper regions of the spectrum (two states with $E^\mathrm{HO}_{n_\rho l n_z}=-21.7E_R$ and two states with $E^\mathrm{HO}_{n_\rho l n_z}=-0.5E_R$). Figure~\ref{Fig:HOPTWKB} specifies the quantum numbers $n_\rho$, $l$, and $n_z$ for the states considered. This figure shows the progression of the energies with respect to expansion order and perturbation order. Comparing the states, we see that the perturbative corrections are larger, in absolute terms, for states higher in the spectrum. Moreover, on the scale of the individual plots, states higher in the spectrum also exhibit slower convergence. Unfortunately, we cannot, in general terms, declare a satisfactory expansion order or perturbation order, as this depends on the specific problem and the desired accuracy. In any case, Figure~\ref{Fig:HOPTWKB} provides sufficient impetus for pursuing an alternative means of describing the motional energy spectrum of atoms trapped in an optical lattice.

In the following two sections, we outline a non-perturbative model that fully respects the anharmonicity and inseparability inherent in the optical lattice potential, Eq.~(\ref{Eq:potential}). Although our model invokes its own approximations, when the harmonic oscillator potential, Eq.~(\ref{Eq:HOpot}), is substituted in place of the full potential, the exact harmonic oscillator energy spectrum is recovered. Relative to the perturbative approach discussed above, our model is a more direct and complete approach for accounting for the differences between these distinct potentials.

\section{The Born-Oppenheimer Approximation}
\label{Sec:BO}

The Born-Oppenheimer (BO) approximation is a ubiquitous tool in molecular theory. It's motivated by a simple principle: the electronic and nuclear dynamics in a molecule evolve on much different timescales due to a large disparity in masses. We identify an analogous principle in the present problem: the axial and radial dynamics of the trapped atom evolve on much different timescales due to a large disparity in optical forces (recalling $\kappa/k\ll1$). We are thus prompted to invoke the BO approximation with the radial and axial degrees of freedom taking the respective roles of the nuclear and electronic degrees of freedom of the molecular problem \cite{myfootnote}.

The first step of the BO approximation is to regard the atom as fixed at a radial distance $\rho$ and to solve for the corresponding axial motion. That is, we solve the eigenvalue equation
\begin{equation}
\left[-\frac{\hbar^2}{2m}\frac{\partial^2}{\partial z^2}+U(\rho,z)\right]\mathcal{Z}_{n_z}(\rho,z)=U_{n_z}(\rho)\mathcal{Z}_{n_z}(\rho,z),
\label{Eq:axialBO}
\end{equation}
where both the eigenfunctions $\mathcal{Z}_{n_z}(\rho,z)$ and the eigenvalues $U_{n_z}(\rho)$ depend on the radial distance $\rho$. The solutions are enumerated by non-negative integer $n_z$, and we take the eigenfunctions to be normalized to unity (see Appendix~\ref{Sec:BOnorm}). The eigenvalues $U_{n_z}(\rho)$, mapped out over $\rho$, are analogous to potential energy curves of the molecular problem.

Next, we treat $U_{n_z}(\rho)$ as a potential governing the radial motion. That is, we solve the eigenvalue equation
\begin{equation}
\begin{gathered}
\left[-\frac{\hbar^2}{2m}\frac{d^2}{d \rho^2}+U_{n_z}(\rho)+\frac{\hbar^2}{2m}\frac{l^2-1/4}{\rho^2}\right]\mathcal{R}_{n_\rho l n_z}(\rho)
\\
=E_{n_\rho l n_z}\,\mathcal{R}_{n_\rho l n_z}(\rho),
\end{gathered}
\label{Eq:radialBO}
\end{equation}
where, as before, the integer $l$ specifies the $z$-component of angular momentum in units of $\hbar$. The last term in square brackets is a centrifugal potential supplementing the radial potential $U_{n_z}(\rho)$. For a given $n_z$ and $l$, the solutions are enumerated by non-negative integer $n_\rho$. 

Finally, we introduce the wave functions
\begin{equation}
\Psi_{n_\rho l n_z}(\rho,\varphi,z)=\frac{1}{\sqrt{2\pi\rho}}\mathcal{R}_{n_\rho l n_z}(\rho)\mathcal{Z}_{n_z}(\rho,z)e^{il\varphi}.
\label{Eq:BOwf}
\end{equation}
Anticipating
\begin{equation*}
\frac{\partial^2}{\partial\rho^2}\mathcal{R}_{n_\rho l n_z}(\rho)\mathcal{Z}_{n_z}(\rho,z)\approx
\mathcal{Z}_{n_z}(\rho,z)
\frac{d^2}{d\rho^2}\mathcal{R}_{n_\rho l n_z}(\rho),
\end{equation*}
which is the mathematical premise of our approximation, it follows that
\begin{equation*}
\left[-\frac{\hbar^2}{2m}\nabla^2+U(\rho,z)-E_{n_\rho l n_z}\right]\Psi_{n_\rho l n_z}(\rho,\varphi,z)
\approx 0.
\end{equation*}
That is, the $E_{n_\rho l n_z}$ and $\Psi_{n_\rho l n_z}(\rho,\varphi,z)$ obtained by the above prescription represent approximate solutions to the time-independent Schr\"{o}dinger equation.

At this point, we must be more concrete regarding our exclusion of tunneling. In the harmonic oscillator approximation, the atoms are effectively confined to the lattice site at the origin (i.e., $|z|<\pi/2k$). For example, for $D=50E_R$ and states with $E^\mathrm{HO}_{n_\rho l n_z}<0$, the harmonic oscillator wave functions (Appendix~\ref{Sec:HOwf}) give a probability of $<\!3\times10^{-5}$ for the atom to be in the region $|z|>\pi/2k$. In similar spirit, here we confine the atom to the lattice site at the origin by supplementing $U(\rho,z)$ in Eq.~(\ref{Eq:potential}) with infinite potential barriers for $|z|>\pi/2k$. In both cases, the lattice site at the origin is understood to be representative of all lattice sites.

The axial eigenvalue equation, Eq.~(\ref{Eq:axialBO}), has analytical solutions in terms of Mathieu functions and characteristic values of the Mathieu functions, as presented in Appendix \ref{Sec:Zsol}. Figure~\ref{Fig:Ucurves} displays the eigenvalues $U_{n_z}(\rho)$ versus $\rho$ for a potential depth $D=50E_R$. These radial potentials increase smoothly and monotonically with $\rho$. The general picture of Fig.~\ref{Fig:Ucurves} holds for other depths, but with generally more (less) curves in the negative energy region for deeper (shallower) traps.

\begin{figure}[tb]
\includegraphics[width=246pt]{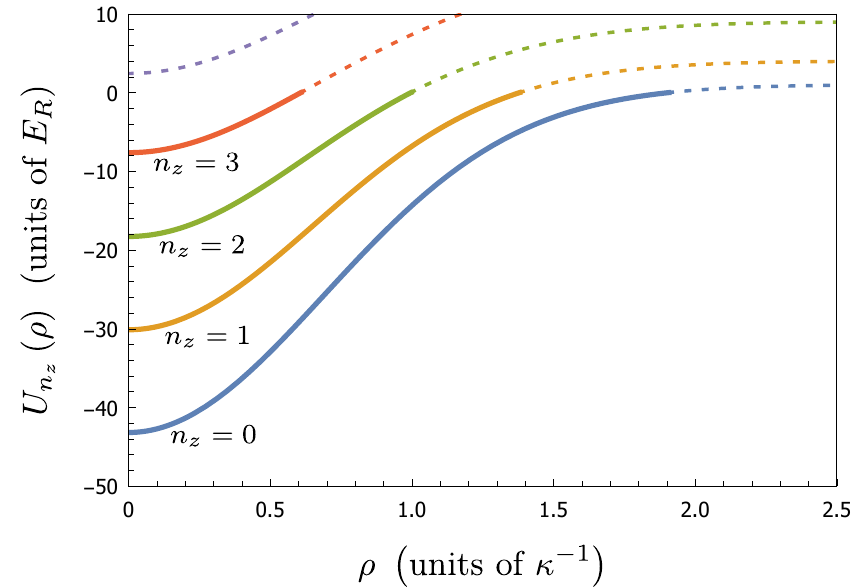}
\caption{Radial potentials $U_{n_z}(\rho)$ for $D=50E_R$. The radial potentials are presented as full curves in the negative energy region relevant to our analysis and as dotted curves otherwise. These are analogous to potential energy curves of the molecular problem.}
\label{Fig:Ucurves}
\end{figure}

Unlike the axial eigenvalue equation, an analytical solution is not available for the subsequent radial eigenvalue equation, Eq.~(\ref{Eq:radialBO}). Consequently, another layer of approximation is required to describe the energy spectrum. We discuss our approach in the following section.

\section{The WKB approximation}
\label{Sec:WKB}

Given our assumptions, namely $\kappa/k\ll1$, the radial potentials $U_{n_z}(\rho)$ increase gradually with $\rho$, and the spectrum of Eq.~(\ref{Eq:radialBO}) will be dense in both $n_\rho$ and $l$. This motivates a WKB (semi-classical) approximation. The familiar one-dimensional WKB approximation is applicable to two-dimensional problems of radial symmetry, so long as the $l^2-1/4$ appearing in the centrifugal potential (refer to Eq.~(\ref{Eq:radialBO})) is replaced with $l^2$ to properly account for behavior of the wave function at the origin~\cite{BerOzo73}.

For a given $n_z$ and $l$, we introduce the WKB phase as a function of energy $E<0$,
\begin{equation}
\phi_{l n_z}(E)=\sqrt{\frac{2m}{\hbar^2}}\int_\mathbb{R}\sqrt{E-U_{n_z}(\rho)-\frac{\hbar^2}{2m}\frac{l^2}{\rho^2}}\,d\rho.
\label{Eq:WKBphase}
\end{equation}
Here and in integrals to follow, the subscript $\mathbb{R}$ on the integration symbol denotes restriction to the region for which the integrand is real. According to the WKB approximation, the energies $E_{n_\rho l n_z}$ are associated with the condition
\begin{equation}
\phi_{ln_z}\left(E_{n_\rho l n_z}\right)=\pi\left(n_\rho+\frac{1}{2}\right).
\label{Eq:WKBcriteria}
\end{equation}
These energies can be numerically evaluated with negligible error. Thus, we have outlined a means to approximate the motional energy spectrum of atoms trapped in an optical lattice.

As an example, we apply the BO+WKB procedure described above to the states considered in Figure~\ref{Fig:HOPTWKB}. The resulting energies are plotted as horizontal dashed lines in Figure~\ref{Fig:HOPTWKB} and can be directly compared to results of the perturbative approach. Generally speaking, we observe that as more perturbation and expansion orders are included in the perturbative approach, the energies tend toward the BO+WKB results.

\begin{figure}[tb]
\includegraphics[width=246pt]{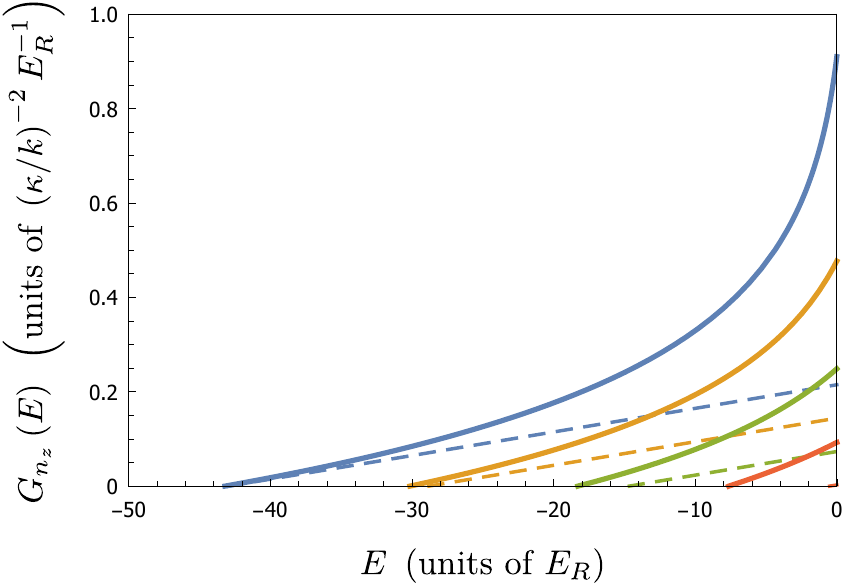}
\caption{Density of states $G_{n_z}(E)$ obtained from the BO+WKB approximation for $D=50E_R$ (full curves). Colors distinguishing different $n_z$ are as in Fig.~\ref{Fig:Ucurves}. The density of states for the harmonic oscillator potential, Eq.~(\ref{Eq:HODOS}), is also shown for $D=50E_R$ (dashed lines).}
\label{Fig:Gcurves}
\end{figure}

Since the energy spectrum is dense in both $n_\rho$ and $l$, we may express it as a density of states for a given $n_z$. We start with the density of states for a given $n_z$ and $l$, which we infer from Eq.~(\ref{Eq:WKBcriteria}) to be equal to $\pi^{-1}\phi_{ln_z}^\prime(E)$. Here and throughout, primes are used to denote derivatives when attached to functions with a single explicit variable. From Eq.~(\ref{Eq:WKBphase}),
\begin{equation}
\phi_{ln_z}^\prime(E)=\frac{1}{2}\sqrt{\frac{2m}{\hbar^2}}\int_\mathbb{R}\frac{1}{\displaystyle\sqrt{E-U_{n_z}(\rho)-\frac{\hbar^2}{2m}\frac{l^2}{\rho^2}}}d\rho.
\label{Eq:phiprime}
\end{equation}
The density of states for a given $n_z$ (inclusive of all $l$) is therefore
\begin{equation*}
G_{n_z}(E)=\frac{1}{2\pi}\sqrt{\frac{2m}{\hbar^2}}\iint_\mathbb{R}\frac{1}{\displaystyle\sqrt{E-U_{n_z}(\rho)-\frac{\hbar^2}{2m}\frac{l^2}{\rho^2}}}d\rho\,dl,
\end{equation*}
where we exploit the fact that the spectrum is dense in $l$ to replace a summation over $l$ with an integration over $l$. Choosing to integrate with respect to $l$ first (see Appendix \ref{Sec:lint}), we obtain
\begin{equation*}
G_{n_z}(E)=\frac{1}{2}\frac{2m}{\hbar^2}\int_0^{R_{n_z}(E)}\rho\,d\rho,
\end{equation*}
where explicit limits are given for the remaining integral. Here $R_{n_z}(E)$ is the inverse function of $U_{n_z}(\rho)$. That is, $U_{n_z}\left(R_{n_z}(E)\right)=E$ and $R_{n_z}(U_{n_z}(\rho))=\rho$. The integral with respect to $\rho$ is trivial, yielding the expression
\begin{equation}
G_{n_z}(E)=\frac{1}{4}\frac{2m}{\hbar^2}\left[R_{n_z}(E)\right]^2.
\label{Eq:DOS}
\end{equation}
In Figure~\ref{Fig:Gcurves}, we present the density of states $G_{n_z}(E)$ resulting from our BO+WKB approximation for the case $D=50E_R$.

\section{The harmonic oscillator potential revisited}
\label{Sec:HOrevisited}

As briefly mentioned at the end of Sec.~\ref{Sec:HOapprox}, if the harmonic oscillator potential, Eq.~(\ref{Eq:HOpot}), is substituted in place of the full potential, Eq.~(\ref{Eq:potential}), the BO+WKB approximation recovers the exact harmonic oscillator energies, Eq.~(\ref{Eq:HOenergies}). This is because the BO approximation is exact if the potential $U(\rho,z)$ is separable in $\rho$ and $z$, with the radial potentials $U_{n_z}(\rho)$ subsequently being equal to the radial part of $U(\rho,z)$ plus an $n_z$-dependent offset, while the WKB approximation is exact if the radial potentials $U_{n_z}(\rho)$ are radial harmonic oscillator potentials with arbitrary offsets.

Since the BO+WKB approximation is exact for the harmonic oscillator potential, we can exploit this potential to check our expression for the density of states, Eq.~(\ref{Eq:DOS}). For $U(\rho,z)\rightarrow U^\mathrm{HO}(\rho,z)$ we readily find
\begin{equation*}
U_{n_z}(\rho)\rightarrow-D+D\kappa^2\rho^2+2\sqrt{DE_R}\left(n_z+\frac{1}{2}\right).
\end{equation*}
From this it follows that
\begin{equation*}
\left[R_{n_z}(E)\right]^2\rightarrow\frac{1}{D\kappa^2}\left[E+D-2\sqrt{DE_R}\left(n_z+\frac{1}{2}\right)\right].
\end{equation*}
Inserting this into the right-hand-side of Eq.~(\ref{Eq:DOS}), we recover the density of states $G^\mathrm{HO}_{n_z}(E)$ given in Eq.~(\ref{Eq:HODOS}), which was derived directly from the spectrum $E^\mathrm{HO}_{n_\rho ln_z}$.

In Figure \ref{Fig:Gcurves}, we compare the density of states for the full potential, given by the BO+WKB model, with the density of states for the harmonic oscillator potential for the case $D=50E_R$. We see that the two cases agree well for the most deeply bound $n_z=0$ states but are largely discrepant otherwise. With the expectation that the BO+WKB model provides a better representation of the density of states of the full potential, this picture gives further reason to be skeptical of the harmonic oscillator energies and wave functions as a zeroth order representation for all but the most deeply bound states.

\section{Lattice light shifts in optical lattice clocks}
\label{Sec:BO+WKBlightshifts}

In an optical lattice clock, the lattice lasers are nominally operated at the magic frequency, defined such that the $E1$ polarizability $\alpha_{E1}$ of the ground ($g$) and excited ($e$) clock states are identical. Consequently, the optical lattice potential is independent of the clock state, and all ``carrier'' transitions $\left|g;n_\rho ln_z\right\rangle\rightarrow\left|e;n_\rho ln_z\right\rangle$ contributing to the spectroscopic signal have a transition frequency equal to the bare atomic transition frequency, $\left|g\right\rangle\rightarrow\left|e\right\rangle$. For high accuracy, however, various ``non-magic'' effects must be considered. These include i) deviation from the magic frequency, leading to a small difference in $E1$ polarizabilities, ii) second order magnetic dipole ($M1$) and electric quadrupole ($E2$) coupling to the lattice field, encapsulated by the $M1$+$E2$ polarizability $\alpha_{M1+E2}$, and iii) fourth order electric dipole coupling to the lattice field, encapsulated by the hyperpolarizability $\beta$. Accounting for these non-magic effects, the potentials for the two clock states generally differ, and the carrier transitions $\left|g;n_\rho ln_z\right\rangle\rightarrow\left|e;n_\rho ln_z\right\rangle$ subsequently include a change in motional energy of the atom, $\Delta E_{n_\rho ln_z}$, in addition to the change in the atom's internal energy. We assume the corresponding clock frequency shift is given by $\Delta E/h$, where $h$ is Planck's constant and $\Delta E$ is the ensemble-average of the motional energy differences.

The non-magic effects lead to small corrections to the potential of each clock state, which can be satisfactorily accounted for with first order perturbation theory. For a Yb optical lattice clock operating within 10 MHz of the magic frequency and at depths below 1000$E_R$, for example, the corrections to the potential of each clock state remain below $10^{-6}D$~\cite{BroPhiBel17}. We take the residual difference in potentials between the clock states to be
\begin{equation*}
\begin{aligned}
\Delta U(\rho,z)=&
-\left(\frac{\mathcal{E}_0}{2}\right)^2\Delta\alpha_{E1}
\,e^{-\kappa^2\rho^2}\cos^2(kz)
\\&
-\left(\frac{\mathcal{E}_0}{2}\right)^2\Delta\alpha_{M1+E2}
\,e^{-\kappa^2\rho^2}\sin^2(kz)
\\&
-\left(\frac{\mathcal{E}_0}{2}\right)^4\Delta\beta
\,e^{-2\kappa^2\rho^2}\cos^4(kz),
\end{aligned}
\end{equation*}
where $\Delta\alpha_{E1}$, $\Delta\alpha_{M1+E2}$, and $\Delta\beta$ are differential atomic parameters between the clock states. In the vicinity of the magic frequency, $\Delta\alpha_{M1+E2}$ and $\Delta\beta$ have negligible dependence on the lattice frequency, while $\Delta\alpha_{E1}$ can be decomposed in terms of a slope and a zero-crossing with respect to the lattice frequency. By definition, the zero-crossing corresponds to the magic frequency. We take this decomposition of $\Delta\alpha_{E1}$ to be implicit throughout.

In terms of the depth, the residual difference in potentials is
\begin{equation}
\begin{aligned}
\Delta U(\rho,z)=&
-D\,\frac{\Delta\alpha_{E1}}{\alpha_{E1}}
\,e^{-\kappa^2\rho^2}\cos^2(kz)
\\&
-D\,\frac{\Delta\alpha_{M1+E2}}{\alpha_{E1}}
\,e^{-\kappa^2\rho^2}\sin^2(kz)
\\&
-D^2\,\frac{\Delta\beta}{\alpha_{E1}^2}
\,e^{-2\kappa^2\rho^2}\cos^4(kz).
\end{aligned}
\label{Eq:dU}
\end{equation}
In the context of our BO+WKB approximation, $\Delta U(\rho,z)$ leads to a residual difference in the radial potentials given by
\begin{equation}
\Delta U_{n_z}(\rho)
=\int_{-\pi/2k}^{+\pi/2k}\left|\mathcal{Z}_{n_z}(\rho,z)\right|^2\Delta U(\rho,z)dz.
\label{Eq:dUnz}
\end{equation}
This in turn leads to a residual difference in the WKB phases given by
\begin{equation}
\Delta\phi_{ln_z}(E)=
-\frac{1}{2}\sqrt{\frac{2m}{\hbar^2}}
\int_\mathbb{R}\frac{\Delta U_{n_z}(\rho)}{\displaystyle\sqrt{E-U_{n_z}(\rho)-\frac{\hbar^2}{2m}\frac{l^2}{\rho^2}}}d\rho,
\label{Eq:Deltaphi}
\end{equation}
which leads to a residual difference in motional energies given by
\begin{equation}
\Delta E_{n_\rho ln_z}=
-\frac{\Delta\phi_{ln_z}(E_{n_\rho ln_z})}{\phi^\prime_{ln_z}(E_{n_\rho ln_z})}.
\label{Eq:dEBOWKB}
\end{equation}
We choose to express $\Delta E_{n_\rho ln_z}$ in the form
\begin{equation}
\begin{aligned}
\Delta E_{n_\rho ln_z}=&
-D\,\frac{\Delta\alpha_{E1}}{\alpha_{E1}}
\,X_{n_\rho ln_z}
-D\,\frac{\Delta\alpha_{M1+E2}}{\alpha_{E1}}
\,Y_{n_\rho ln_z}
\\&
-D^2\,\frac{\Delta\beta}{\alpha_{E1}^2}
\,Z_{n_\rho ln_z}.
\end{aligned}
\label{Eq:dEXYZstate}
\end{equation}
By comparison with Eq.~(\ref{Eq:dU}), it is evident that the dimensionless factors $X_{n_\rho ln_z}$, $Y_{n_\rho ln_z}$, and $Z_{n_\rho ln_z}$ are restricted to the range $[0,1]$. For an atom residing precisely at the origin (i.e., at the center of the lattice site), these factors would be $X_{n_\rho ln_z}=1$, $Y_{n_\rho ln_z}=0$, and $Z_{n_\rho ln_z}=1$. The spread of the motional state wave function beyond the origin leads to a deviation from these ``nominal'' values. Finally, we write
\begin{equation}
\Delta E=
-D\,\frac{\Delta\alpha_{E1}}{\alpha_{E1}}
\,X
-D\,\frac{\Delta\alpha_{M1+E2}}{\alpha_{E1}}
\,Y
-D^2\,\frac{\Delta\beta}{\alpha_{E1}^2}
\,Z,
\label{Eq:dEXYZ}
\end{equation}
with dimensionless factors $X$, $Y$, and $Z$ being the ensemble-averages of $X_{n_\rho ln_z}$, $Y_{n_\rho ln_z}$, and $Z_{n_\rho ln_z}$. These factors are likewise restricted to the range $[0,1]$.

From Eqs.~(\ref{Eq:dU}) through (\ref{Eq:dEXYZstate}) and Eq.~(\ref{Eq:phiprime}), we find the following expression for $X_{n_\rho ln_z}$,
\begin{equation*}
X_{n_\rho ln_z}=
\frac{\displaystyle\int_\mathbb{R}
\frac{x_{n_z}(\rho)}{\displaystyle\sqrt{E_{n_\rho ln_z}-U_{n_z}(\rho)-\frac{\hbar^2}{2m}\frac{l^2}{\rho^2}}}d\rho}
{\displaystyle\int_\mathbb{R}\frac{1}{\displaystyle\sqrt{E_{n_\rho ln_z}-U_{n_z}(\rho)-\frac{\hbar^2}{2m}\frac{l^2}{\rho^2}}}d\rho},
\end{equation*}
where the dimensionless function $x_{n_z}(\rho)$ reads
\begin{equation*}
x_{n_z}(\rho)
=e^{-\kappa^2\rho^2}\int_{-\pi/2k}^{+\pi/2k}\left|\mathcal{Z}_{n_z}(\rho,z)\right|^2\cos^2(kz)dz.
\end{equation*}
Analogous expressions hold for $Y_{n_\rho ln_z}$ and $Z_{n_\rho ln_z}$, with $x_{n_z}(\rho)$ being replaced with 
\begin{gather*}
y_{n_z}(\rho)
=e^{-\kappa^2\rho^2}\int_{-\pi/2k}^{+\pi/2k}\left|\mathcal{Z}_{n_z}(\rho,z)\right|^2\sin^2(kz)dz,
\\
z_{n_z}(\rho)
=e^{-2\kappa^2\rho^2}\int_{-\pi/2k}^{+\pi/2k}\left|\mathcal{Z}_{n_z}(\rho,z)\right|^2\cos^4(kz)dz.
\end{gather*}
We note the relation $x_{n_z}(\rho)+y_{n_z}(\rho)=e^{-\kappa^2\rho^2}$. These expressions can be used to numerically evaluate $X_{n_\rho ln_z}$, $Y_{n_\rho ln_z}$, and $Z_{n_\rho ln_z}$ for specific motional states.

To evaluate the ensemble-average factors $X$, $Y$, and $Z$, a distribution over motional states must be specified. To demonstrate such an evaluation, we initially suppose a Boltzmann distribution. It follows that $X$ is given by
\begin{equation}
X=
\frac{\displaystyle\sum_{n_\rho ln_z}X_{n_\rho ln_z}e^{-E_{n_\rho ln_z}/k_BT}}
{\displaystyle\sum_{n_\rho ln_z}e^{-E_{n_\rho ln_z}/k_BT}},
\label{Eq:dEboltz}
\end{equation}
where $k_B$ is Boltzmann's constant, $T$ is the temperature, and the summations run over all states $E_{n_\rho ln_z}<0$. We exploit the fact that the spectrum is dense in $n_\rho$ and $l$ to replace the summations over $n_\rho$ and $l$ in Eq.~(\ref{Eq:dEboltz}) with integrals. For the integration over $n_\rho$, we subsequently make a change of variable from $n_\rho$ to energy $E$, with $E_{n_\rho ln_z}\rightarrow E$. This change of variable must also incorporate the density of states for a given $n_z$ and $l$, such that $dn_\rho\rightarrow\pi^{-1}\phi^\prime_{ln_z}(E)dE$. It follows that
\begin{equation*}
X=
\frac{\displaystyle\sum_{n_z}\iiint_\mathbb{R}
\frac{x_{n_z}(\rho)\,e^{-E/k_BT}}{\displaystyle\sqrt{E-U_{n_z}(\rho)-\frac{\hbar^2}{2m}\frac{l^2}{\rho^2}}}d\rho\,dl\,dE}
{\displaystyle\sum_{n_z}\iiint_\mathbb{R}\frac{e^{-E/k_BT}}{\displaystyle\sqrt{E-U_{n_z}(\rho)-\frac{\hbar^2}{2m}\frac{l^2}{\rho^2}}}d\rho\,dl\,dE}.
\end{equation*}
Performing the integrals with respect to $l$ first (Appendix~\ref{Sec:lint}), this becomes
\begin{equation*}
X=
\frac{\displaystyle\sum_{n_z}\int_{0}^{R_{n_z}(0)}\int_{U_{n_z}(\rho)}^{0}x_{n_z}(\rho)\,\rho\,e^{-E/k_BT}
dE\,d\rho}
{\displaystyle\sum_{n_z}\int_{0}^{R_{n_z}(0)}\int_{U_{n_z}(\rho)}^{0}\rho\,e^{-E/k_BT}dE\,d\rho},
\end{equation*}
where we have chosen an order for the remaining integrals and provided explicit limits. Subsequently performing the integrals with respect to $E$, we get
\begin{equation}
X=
\frac{\displaystyle\sum_{n_z}\int_{0}^{R_{n_z}(0)}x_{n_z}(\rho)\,\rho\left(e^{-U_{n_z}(\rho)/k_BT}-1\right)d\rho}
{\displaystyle\sum_{n_z}\int_{0}^{R_{n_z}(0)}\rho\left(e^{-U_{n_z}(\rho)/k_BT}-1\right)d\rho}.
\label{Eq:Xexpression}
\end{equation}
The remaining integrals with respect to $\rho$ can be performed numerically. In the case of the numerator, an analytical solution is available, as demonstrated in Appendix~\ref{Sec:ananum}. Expressions analogous to Eq.~(\ref{Eq:Xexpression}) hold for $Y$ and $Z$.

\begin{figure*}[tb]
\subfloat{\includegraphics[width=160pt]{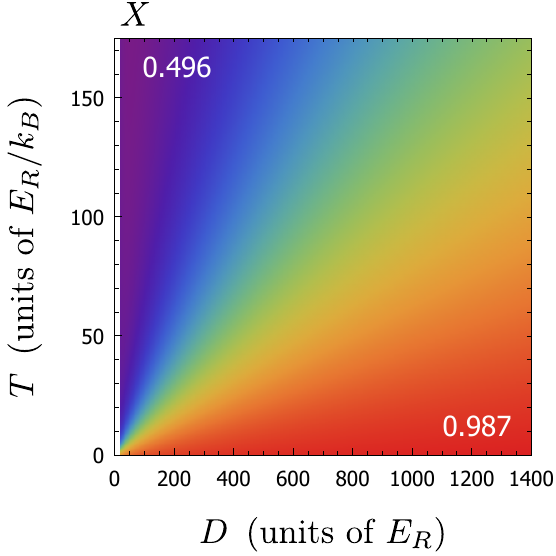}}%
\hspace{15pt}%
\subfloat{\includegraphics[width=160pt]{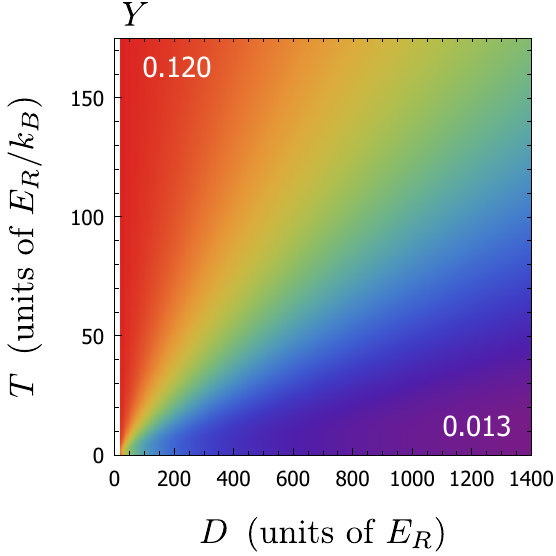}}%
\hspace{15pt}%
\subfloat{\includegraphics[width=160pt]{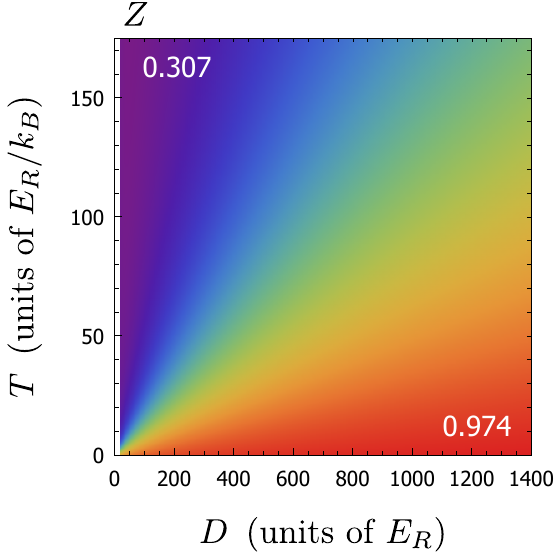}}%
\\
\subfloat{\includegraphics[width=160pt]{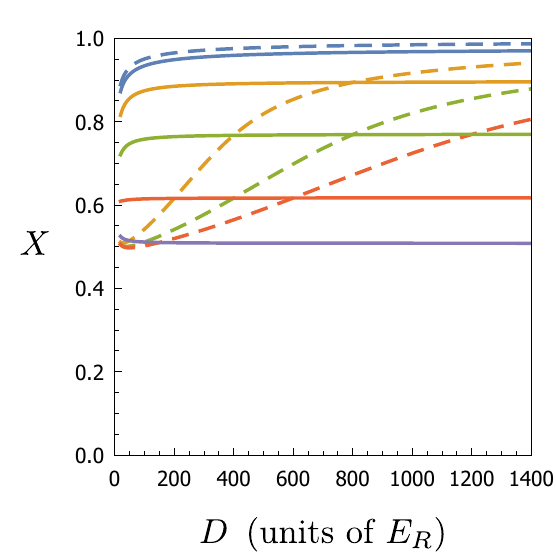}}%
\hspace{15pt}%
\subfloat{\includegraphics[width=160pt]{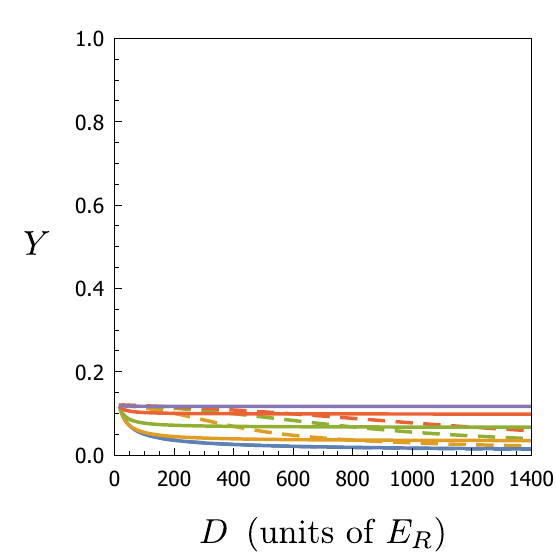}}%
\hspace{15pt}%
\subfloat{\includegraphics[width=160pt]{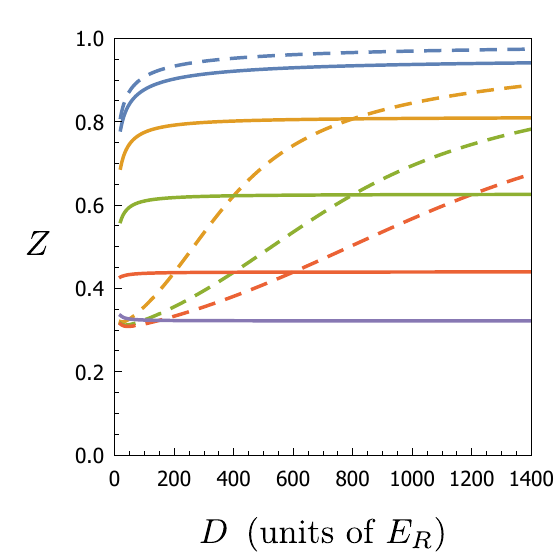}}%
\\
\subfloat{\includegraphics[width=200pt]{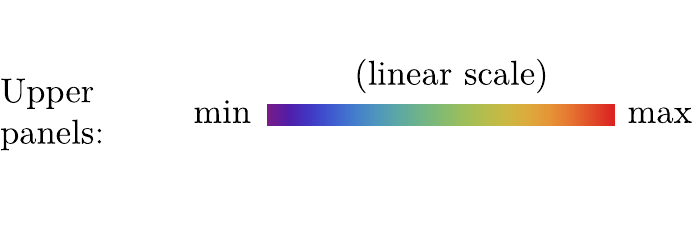}}%
\hspace{50pt}%
\subfloat{\includegraphics[width=200pt]{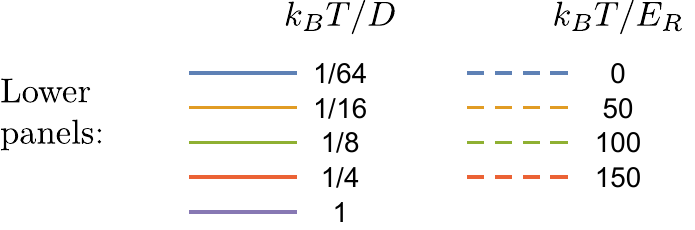}}%
\caption{Factors $X$, $Y$, and $Z$ evaluated with the BO+WKB approximation. A Boltzmann distribution over motional states is assumed. Upper panels: Results for $X$, $Y$, and $Z$ are plotted over a range of depths and temperatures. Maximum and minimum values are given in the corresponding corners of the plots. Lower panels: Results for $X$, $Y$, and $Z$ are plotted versus depth assuming that the temperature is either proportional to depth (full curves) or independent of depth (dashed curves). The minimum depth plotted in all cases is $D=20E_R$. For reference, $E_R/k_B$ equals 0.096 $\mu$K for Yb and 0.17 $\mu$K for Sr.}
\label{Fig:XYZcombined}
\end{figure*}

Figure~\ref{Fig:XYZcombined} presents $X$, $Y$, and $Z$ evaluated over a range of depths and temperatures. We observe that greater depths and lower temperatures yield results closer to the ``nominal'' values (unity for $X$ and $Z$, zero for $Y$). This is the expected behavior, as greater depths and lower temperatures imply a higher degree of atomic localization near the center of the lattice site. Figure~\ref{Fig:XYZcombined} also presents $X$, $Y$, and $Z$ versus depth assuming that the temperature is either i) proportional to the depth or ii) independent of the depth. Results are shown for a number of different proportionality constants and fixed temperatures. The diversity of these curves elucidates an important fact: how the clock shift varies with depth is inescapably linked to how the motional state distribution varies with depth. This is of relevance for optical lattice clocks, because lattice light shifts are typically characterized experimentally by modulating the depth. However, only recently has this fact been shown its due appreciation~\cite{BroPhiBel17,UshTakKat18,NemJorYan19}.

\section{Comparison of lattice light shift models}
\label{Sec:comparemodels}

In this section, we compare lattice light shift models recently employed in Refs.~\cite{BroPhiBel17,UshTakKat18} with the BO+WKB model developed in this work. To this end, some consideration is given to the experimental conditions in Refs.~\cite{BroPhiBel17,UshTakKat18}. We emphasize, however, that the goal here is not to precisely capture details of these experiments, but rather to provide meaningful comparisons of the lattice light shift models under like conditions.

We start by writing $X$ as
\begin{equation*}
X=\sum_{n_\rho ln_z}p_{n_\rho ln_z}X_{n_\rho ln_z},
\end{equation*}
with $p_{n_\rho ln_z}$ being the fractional population for the specific motional state. Next, we partition $X$ into contributions for each $n_z$ as
\begin{equation*}
X=\sum_{n_z}P_{n_z}X_{n_z},
\end{equation*}
where $P_{n_z}$ and $X_{n_z}$ are the fractional population and average value of $X_{n_\rho ln_z}$ for the specific $n_z$. Namely,
\begin{gather*}
P_{n_z}=\sum_{n_\rho l}p_{n_\rho ln_z},
\\
X_{n_z}=\frac{\displaystyle\sum_{n_\rho l}p_{n_\rho ln_z}X_{n_\rho ln_z}}{\displaystyle\sum_{n_\rho l}p_{n_\rho ln_z}}.
\end{gather*}
For each $n_z$, we assume the distribution over states is described by a Boltzmann distribution with radial temperature $T_{\rho}$. That is,
\begin{equation}
X_{n_z}=
\frac{\displaystyle\sum_{n_\rho l}X_{n_\rho l n_z}e^{-E_{n_\rho ln_z}/k_BT_{\rho}}}
{\displaystyle\sum_{n_\rho l}e^{-E_{n_\rho ln_z}/k_BT_{\rho}}},
\label{Eq:XnzTrho}
\end{equation}
where, as before, the summations run over states $E_{n_\rho ln_z}<0$. For the BO+WKB model, an expression for $X_{n_z}$ is given by the right-hand-side of Eq.~(\ref{Eq:Xexpression}) with $T\rightarrow T_{\rho}$ and the specific $n_z$-term isolated from each summation. Namely,
\begin{equation}
X_{n_z}=
\frac{\displaystyle\int_{0}^{R_{n_z}(0)}x_{n_z}(\rho)\,\rho\left(e^{-U_{n_z}(\rho)/k_BT_{\rho}}-1\right)d\rho}
{\displaystyle\int_{0}^{R_{n_z}(0)}\rho\left(e^{-U_{n_z}(\rho)/k_BT_{\rho}}-1\right)d\rho}.
\label{Eq:Xn}
\end{equation}
Analogous expressions hold for $Y$ and $Z$. The factors $X_{n_z}$, $Y_{n_z}$, and $Z_{n_z}$ are restricted to the range $[0,1]$.

The light shift models employed in Brown~{\it et al.}~\cite{BroPhiBel17} and Ushijima~{\it et al.}~\cite{UshTakKat18} both have their roots in the harmonic oscillator approximation. Brown~{\it et al.}\ employed a perturbative approach similar to that described in Sec.~\ref{Sec:HOapprox}, but with non-magic effects incorporated in the potential from the start. Terms fourth order in the series expansion of the coordinates were evaluated at first order in pertubation theory. Taking a difference between the clock states and only retaining contributions first order in the non-magic effects results in a clock shift expression for a specific motional state (Eq.~(2) of Ref.~\cite{BroPhiBel17}). For averaging over the motional states of a given $n_z$, the cross-dimensional correction is excluded from energies in the Boltzmann weighting factor, as are the small non-magic effects. Given $\kappa/k\ll1$, we infer results for $X_{n_z}$, $Y_{n_z}$, and $Z_{n_z}$ in this approach to be
\begin{align*}
X_{n_z}={}&
\left[1-\left(\frac{k_BT_{\rho}}{D}\right)\eta^{(1)}_{n_z}\right]-\left(n_z+\frac{1}{2}\right)
\left(\frac{D}{E_R}\right)^{-1/2},
\\
Y_{n_z}={}&
\left(n_z+\frac{1}{2}\right)\left[1-\frac{1}{2}\left(\frac{k_BT_{\rho}}{D}\right)\eta^{(1)}_{n_z}\right]\left(\frac{D}{E_R}\right)^{-1/2},
\\
Z_{n_z}={}&
\left[1-2\left(\frac{k_BT_{\rho}}{D}\right)\eta^{(1)}_{n_z}+2\left(\frac{k_BT_{\rho}}{D}\right)^2\eta^{(2)}_{n_z}\right]
\\&
-2\left(n_z+\frac{1}{2}\right)\left[1-\left(\frac{k_BT_{\rho}}{D}\right)\eta^{(1)}_{n_z}\right]\left(\frac{D}{E_R}\right)^{-1/2}
\\&
+\frac{3}{2}\left(n_z^2+n_z+\frac{1}{2}\right)\left(\frac{D}{E_R}\right)^{-1}.
\end{align*}
The factors $\eta^{(p)}_{n_z}$ arise from the averaging procedure; they are attributed to i) radial and axial anharmonic corrections to the energies in the Boltzmann weighting factor and ii) a restriction to states with energies less than zero (being representative of trapped motional states). In the limit $k_BT_\rho/D\rightarrow0$, these two considerations are inconsequential, with $\eta^{(p)}_{n_z}\rightarrow1$. More details are provided in Appendix~\ref{Sec:Brownetafac}, where an expression for $\eta^{(p)}_{n_z}$ can be found.

Conceptually, Ushijima~{\it et al.}\ take a different approach. They incorporate non-magic effects in the potential from the start, though initially neglect the radial degrees of freedom (effectively setting $\rho=0$). For the resulting one-dimensional potential, they invoke a familiar clock shift expression for a specific $n_z$ (Eq.~(1) of Ref.~\cite{UshTakKat18}). This expression is derived in the same manner as the clock shift expression from Brown~{\it et al.}, but limited to just the axial degree of freedom~\cite{KatOvsMar15}. Independently, the radial degrees of freedom are treated classically. The Boltzmann distribution is translated into a probability distribution for the radial coordinate $\rho$. This step requires assuming a potential for the radial confinement. For this purpose, the radial part of the harmonic oscillator potential, Eq.~(\ref{Eq:HOpot}), is taken. The radial and axial degrees of freedom are then merged by making the substitution $D\rightarrow De^{-\kappa^2\rho^2}$ in the one-dimensional clock shift expression and averaging over the probability distribution for $\rho$. The effects of this radial averaging are encapsulated by dimensionless ``reduction factors'' $\zeta_j$. We infer results for $X_{n_z}$, $Y_{n_z}$, and $Z_{n_z}$ in this approach to be
\begin{align*}
X_{n_z}={}&
\zeta_1
-\left(n_z+\frac{1}{2}\right)\zeta_{1/2}
\left(\frac{D}{E_R}\right)^{-1/2},
\\
Y_{n_z}={}&
\left(n_z+\frac{1}{2}\right)\zeta_{1/2}
\left(\frac{D}{E_R}\right)^{-1/2},
\\
Z_{n_z}={}&
\zeta_2
-2\left(n_z+\frac{1}{2}\right)\zeta_{3/2}\left(\frac{D}{E_R}\right)^{-1/2}
\\&
+\frac{3}{2}\left(n_z^2+n_z+\frac{1}{2}\right)\zeta_{1}
\left(\frac{D}{E_R}\right)^{-1}.
\end{align*}
Ushijima~{\it et al.}\ give an approximate expression for the reduction factors,
\begin{equation}
\zeta_j
\approx
1-j\left(\frac{k_BT_{\rho}}{D}\right).
\label{Eq:Ushfo}
\end{equation}
Following their procedure, however, a simple analytical solution is available,
\begin{equation}
\zeta_j
=\left[1+j\left(\frac{k_BT_{\rho}}{D}\right)\right]^{-1},
\label{Eq:Ushfull}
\end{equation}
with Eq.~(\ref{Eq:Ushfo}) being correct to first order in $k_BT_{\rho}/D$. Below we consider both cases, with ``modified'' being used to designate the use of Eq.~(\ref{Eq:Ushfull}) in place of Eq.~(\ref{Eq:Ushfo}). In either case, no energy restriction is imposed to exclude states not representative of trapped motional states.

\begin{figure*}[tb]
\subfloat{\includegraphics[width=160pt]{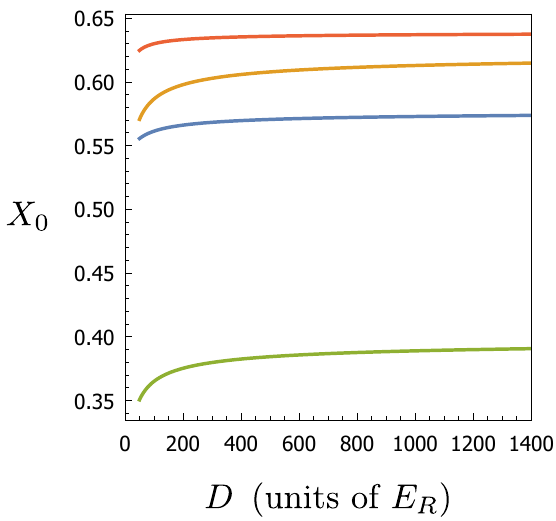}}%
\hspace{15pt}%
\subfloat{\includegraphics[width=160pt]{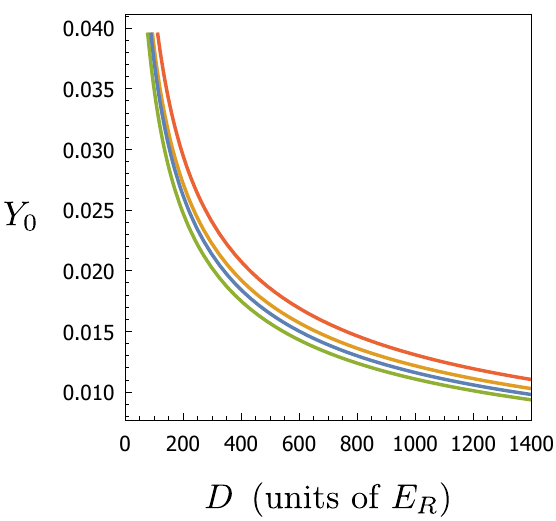}}%
\hspace{15pt}%
\subfloat{\includegraphics[width=160pt]{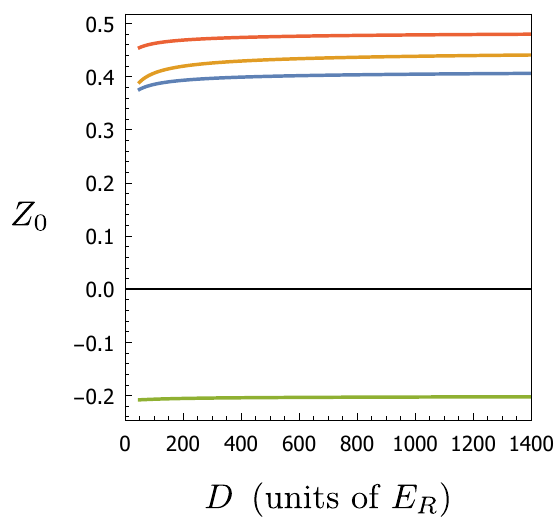}}%
\\
\subfloat{\includegraphics[width=160pt]{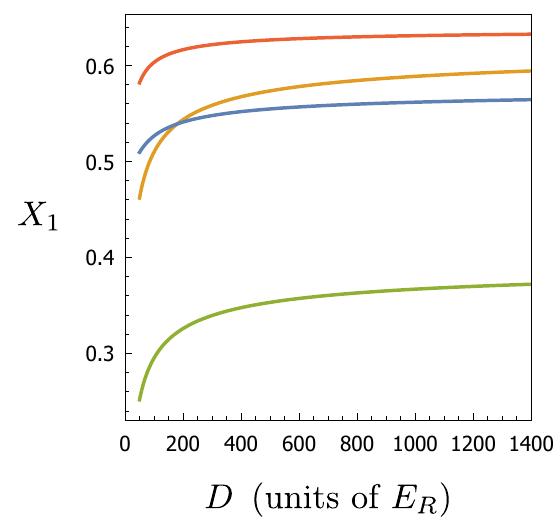}}%
\hspace{15pt}%
\subfloat{\includegraphics[width=160pt]{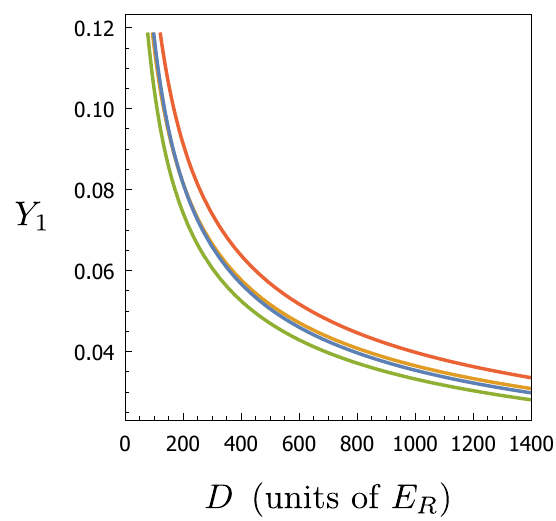}}%
\hspace{15pt}%
\subfloat{\includegraphics[width=160pt]{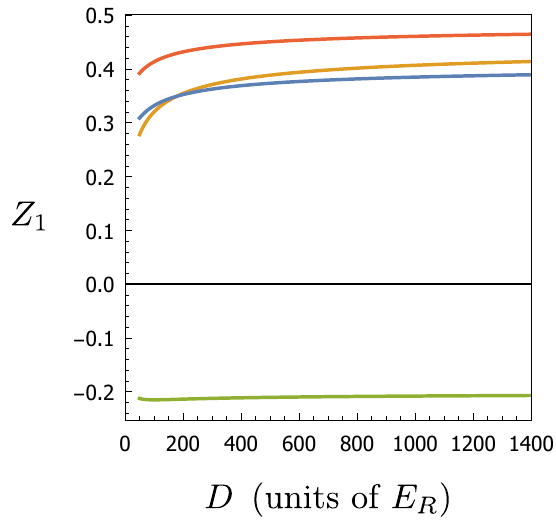}}%
\\
\subfloat{\includegraphics[width=160pt]{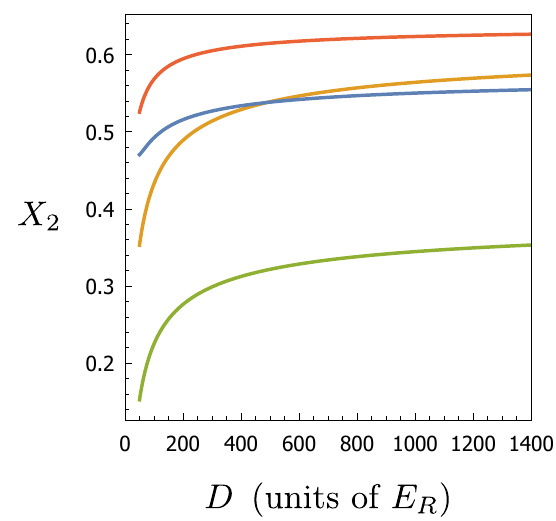}}%
\hspace{15pt}%
\subfloat{\includegraphics[width=160pt]{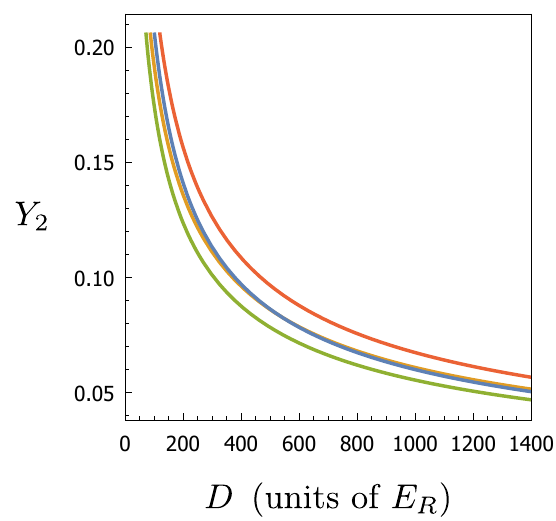}}%
\hspace{15pt}%
\subfloat{\includegraphics[width=160pt]{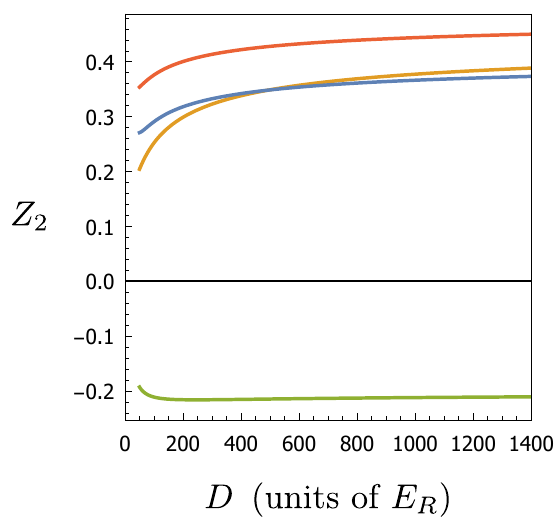}}%
\caption{Factors $X_{n_z}$, $Y_{n_z}$, and $Z_{n_z}$ versus $D$ assuming a radial temperature given by $k_BT_\rho=0.6D$. Top, middle, and bottom rows correspond to $n_z=0$, 1, and 2, respectively. The Brown {\it et al.}\ model (red), Ushijima~{\it et al.}\ model (green), and modified Ushijima~{\it et al.}\ model (yellow) are compared to the BO+WKB model (blue). The lowest depth plotted in each case is $D=50E_R$.}
\label{Fig:XYZcompareBrownconditions}
\end{figure*}

\begin{figure*}[tb]
\subfloat{\includegraphics[width=160pt]{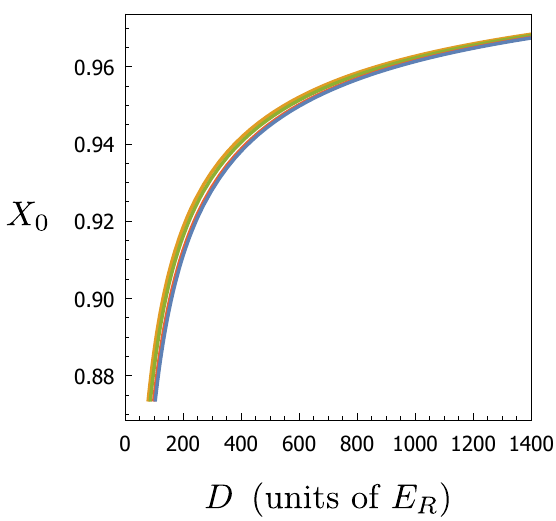}}%
\hspace{15pt}%
\subfloat{\includegraphics[width=160pt]{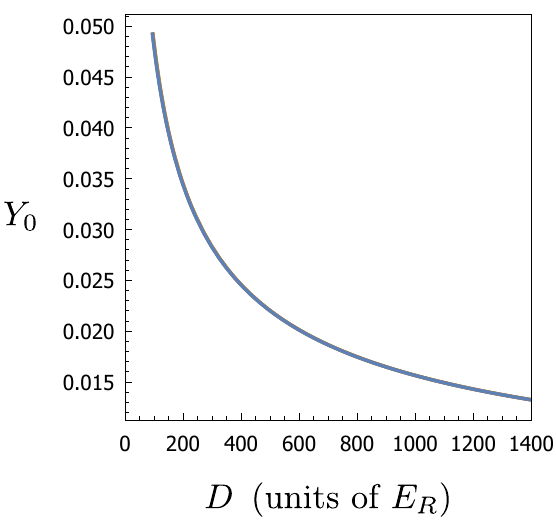}}%
\hspace{15pt}%
\subfloat{\includegraphics[width=160pt]{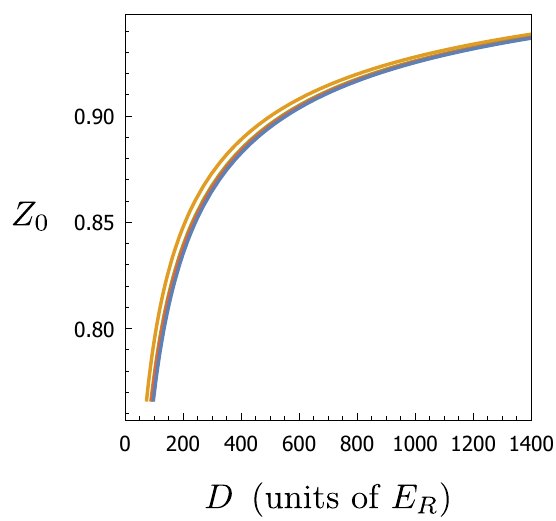}}%
\\
\subfloat{\includegraphics[width=160pt]{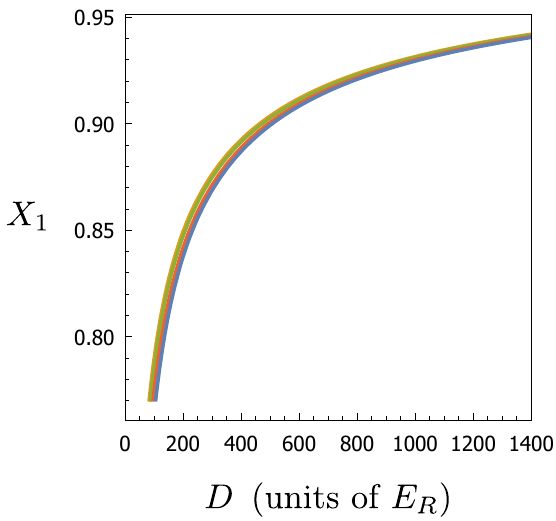}}%
\hspace{15pt}%
\subfloat{\includegraphics[width=160pt]{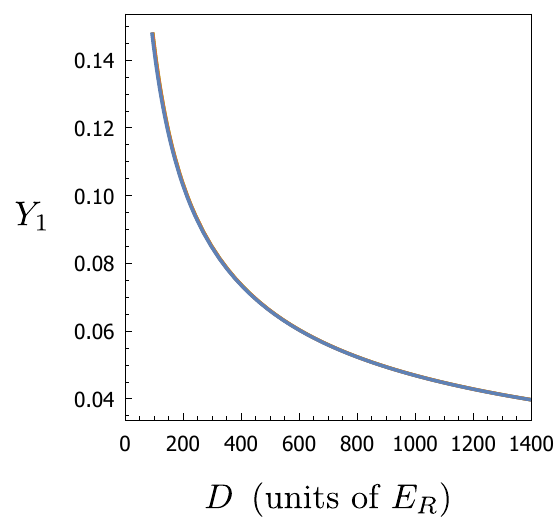}}%
\hspace{15pt}%
\subfloat{\includegraphics[width=160pt]{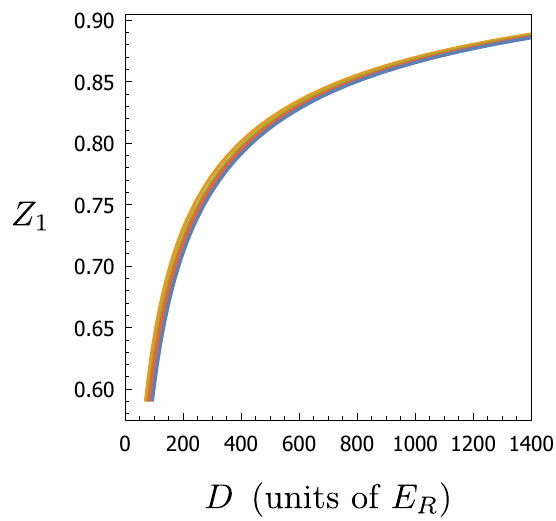}}%
\\
\subfloat{\includegraphics[width=160pt]{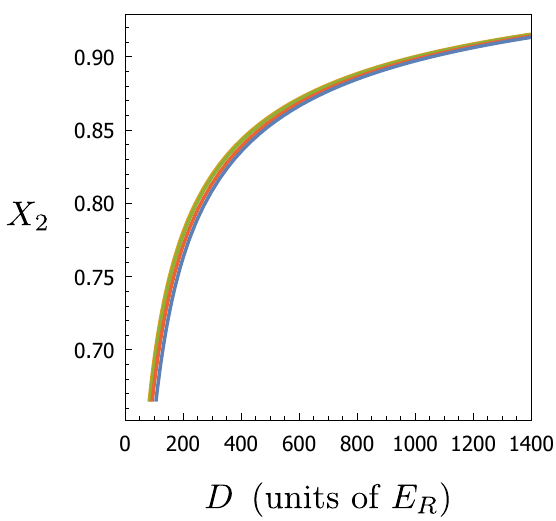}}%
\hspace{15pt}%
\subfloat{\includegraphics[width=160pt]{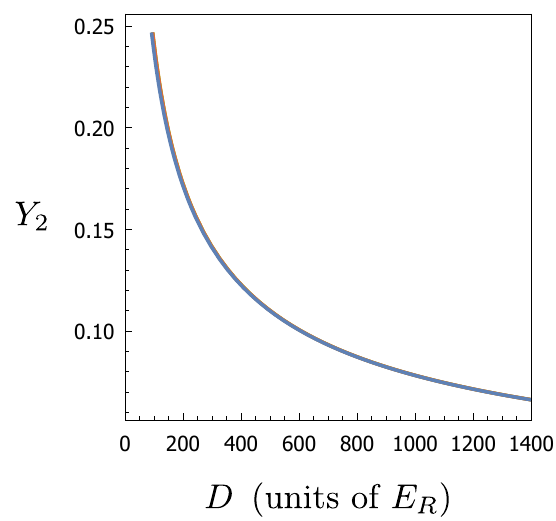}}%
\hspace{15pt}%
\subfloat{\includegraphics[width=160pt]{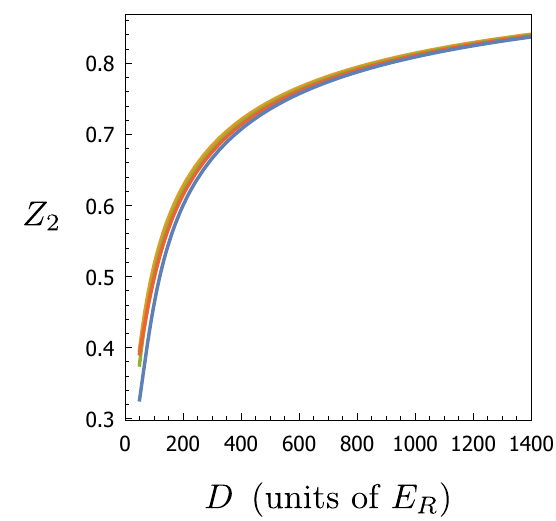}}%
\caption{Factors $X_{n_z}$, $Y_{n_z}$, and $Z_{n_z}$ versus $D$ assuming a radial temperature given by $k_BT_\rho=0.7\sqrt{DE_R}$. Top, middle, and bottom rows correspond to $n_z=0$, 1, and 2, respectively. The Brown {\it et al.}\ model (red), Ushijima~{\it et al.}\ model (green), and modified Ushijima~{\it et al.}\ model (yellow) are compared to the BO+WKB model (blue). The lowest depth plotted in each case is $D=50E_R$.}
\label{Fig:XYZcompareUshiconditions}
\end{figure*}

The Yb optical lattice clock of Brown {\it et al.}\ operated at depths in the range $50E_R\lesssim D\lesssim1400E_R$. Sideband fitting~\cite{BlaThoCam09} suggests an approximate relationship $k_BT_\rho\approx0.6D$ over this range. The Sr optical lattice clock of Ushijima {\it et al.}\ operated by loading the atoms into a lattice at a reference depth of $272E_R$, then adiabatically ramping to a final depth in the range $150E_R\lesssim D\lesssim1150E_R$. Adiabiticity implies an approximate proportionality $T_\rho\propto\sqrt{D}$, while the radial temperature at the reference depth was inferred from time-of-flight thermometry to be $T_\rho\approx2~\mu\text{K}$. This implies an approximate relationship $k_BT_\rho\approx0.7\sqrt{DE_R}$. Figures \ref{Fig:XYZcompareBrownconditions} and \ref{Fig:XYZcompareUshiconditions} compare results for $X_{n_z}$, $Y_{n_z}$, and $Z_{n_z}$ for the different models assuming these two distinct relationships for $T_\rho$ versus $D$. The lowest-$n_z$ cases of $n_z=0$, 1, and 2 are displayed.

In Fig.~\ref{Fig:XYZcompareBrownconditions}, we see that the Brown {\it et al.}, Ushijima {\it et al.}, and modified Ushijima {\it et al.}\ models are all in fair agreement with the BO+WKB model for $Y_{n_z}$. This, however, is not the case for $X_{n_z}$ and $Z_{n_z}$. In particular, the Ushijima {\it et al.}\ model differs substantially from the BO+WKB model and even yields unphysical negative results for $Z_{n_z}$. The Brown {\it et al.}\ and modified Ushijima {\it et al.}\ models exhibit more reasonable agreement with the BO+WKB model. While the modified Ushijima {\it et al.}\ model produces curves numerically closer to the BO+WKB model, the Brown {\it et al.}\ model produces curves with functional behavior closer to the BO+WKB model. In Fig.~\ref{Fig:XYZcompareUshiconditions}, we see that the overall agreement between the models is much better. In fact, curves for the different models are largely indistinguishable from one another on the displayed scale, especially for $Y_{n_z}$. We attribute the better agreement to the lower radial temperature (in units of $E_R/k_B$), implying less population in the higher-lying motional states.

\begin{figure*}[tb]
\subfloat{\includegraphics[width=160pt]{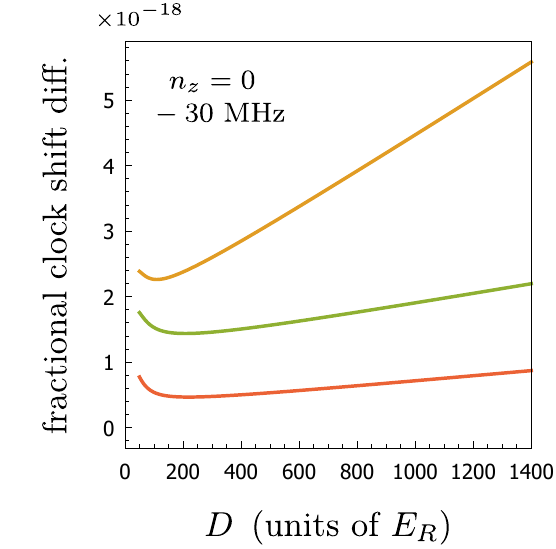}}%
\hspace{15pt}%
\subfloat{\includegraphics[width=160pt]{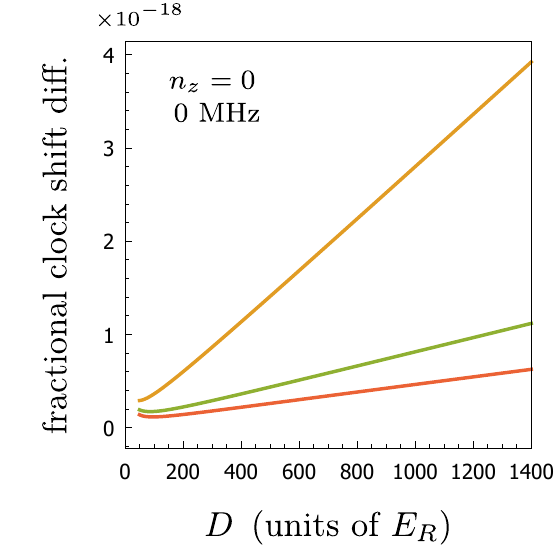}}%
\hspace{15pt}%
\subfloat{\includegraphics[width=160pt]{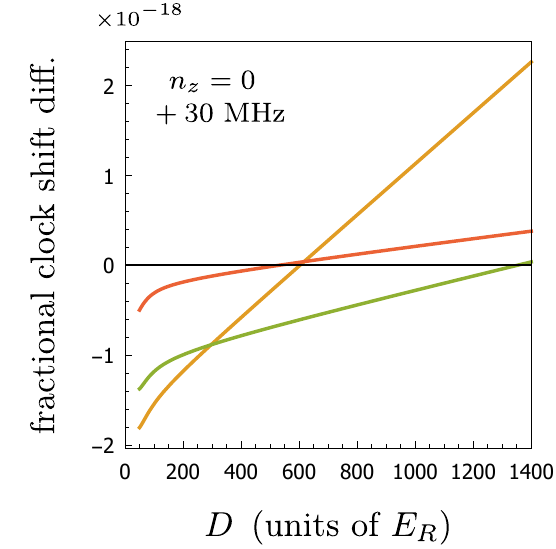}}%
\\
\subfloat{\includegraphics[width=160pt]{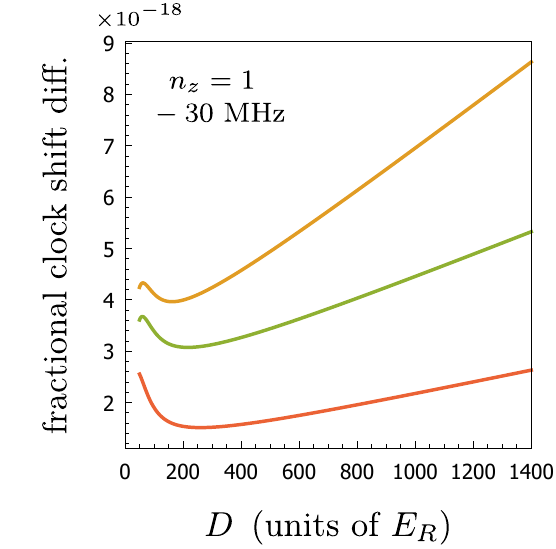}}%
\hspace{15pt}%
\subfloat{\includegraphics[width=160pt]{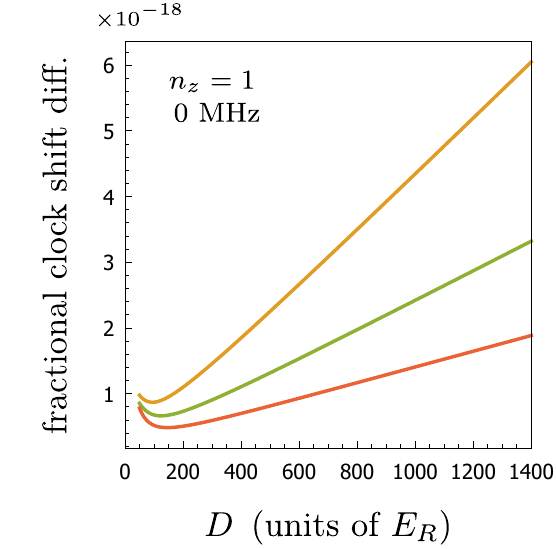}}%
\hspace{15pt}%
\subfloat{\includegraphics[width=160pt]{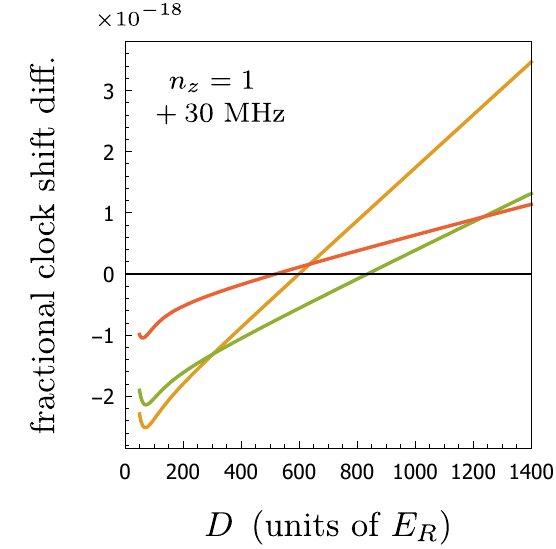}}%
\caption{Difference in the evaluated fractional clock shift of Sr relative to the BO+WKB model, assuming a radial temperature $k_BT_\rho=0.7\sqrt{DE_R}$, $P_{n_z}=\delta_{n_z,0}$ or $P_{n_z}=\delta_{n_z,1}$ (labeled ``$n_z=0$'' and ``$n_z=1$,'' respectively), lattice frequency detunings $-30$ MHz, 0 MHz, or $+30$ MHz relative to the magic frequency (as labeled), and atomic parameters reported in Ref.~\cite{UshTakKat18}. The Brown {\it et al.}\ model (red), Ushijima~{\it et al.}\ model (green), and modified Ushijima~{\it et al.}\ model (yellow) are plotted as differences relative to the BO+WKB model. The lowest depth plotted in each case is $D=50E_R$.}
\label{Fig:shiftdiffUshiconditions}
\end{figure*}

While the overall agreement in Fig.~\ref{Fig:XYZcompareUshiconditions} appears very good, it is interesting to see how the small differences in these curves translate into a difference for the clock shift. To make this translation, we must assume a distribution for $P_{n_z}$, as well as values for the atomic parameters appearing in Eq.~(\ref{Eq:dEXYZ}). Using quenched sideband cooling and subsequent rapid adiabatic passage excitation, Ushijima~{\it et al.}\ prepared atomic ensembles with either $P_{n_z}\approx\delta_{n_z,0}$ or $P_{n_z}\approx\delta_{n_z,1}$. Using their (unmodified) model to interpret lattice light shift measurements, they extracted atomic parameters for Sr. Taking $k_BT_\rho=0.7\sqrt{DE_R}$, $P_{n_z}=\delta_{n_z,0}$ or $P_{n_z}=\delta_{n_z,1}$, and atomic parameters reported by Ushijima~{\it et al.}, we evaluate the fractional clock shift for Sr over a range of depths for three different lattice frequency detunings relative to the magic frequency. Figure~\ref{Fig:shiftdiffUshiconditions} presents results for the Brown {\it et al.}, Ushijima {\it et al.}, and modified Ushijima {\it et al.} models, plotted as a difference relative to the BO+WKB model. We observe that the differences relative to the BO+WKB model, as well as the differences between the models, are on the order of $10^{-18}$. That is, assuming that the BO+WKB model gives the best representation of the clock shift for the assumed conditions, this suggests that the other three models do not support accuracy at the low-$10^{-18}$ level. However, we emphasize that this only concerns absolute accuracy of the models, in which they are used to directly evaluate the clock shift given well-defined conditions and atomic parameters. In the following section, we see that the absolute accuracy of a model can be relaxed if the model is invoked in a different manner, as in Ref.~\cite{BroPhiBel17}.

Before concluding this section, we note that the atomic parameter $\Delta\alpha_{M1+E2}/\alpha_{E1}$ reported by Ushijima {\it et al.}\ differs by tens of $\sigma$, where $\sigma$ represents combined standard deviation, from consistent theoretical values for this parameter~\cite{PorSafSaf18,WuTanShi19,Bel19unpublished}. While Fig.~\ref{Fig:shiftdiffUshiconditions} may suggest some degree of error in the Ushijima {\it et al.}\ model used to extract this and other Sr parameters from measurement, it does not appear to suggest model error significant enough to account for a discrepancy of this size. Thus, more theoretical or experimental investigation will be required to resolve this issue.

\section{Reparametrization of the lattice light shift based on experimental conditions}

The BO+WKB model outlined in this paper was developed in conjunction with the experimental work of Brown {\it et al.}, as potential shortcomings in the pedagogically simpler Brown~{\it et al.}\ model were recognized early. For the Yb optical lattice clock of Brown~{\it et al.}, sideband spectra are consistent with a distribution over motional states characterized by a radial temperature $T_\rho$ and an axial temperature $T_z$~\cite{BlaThoCam09}. If the radial and axial degrees of freedom were perfectly decoupled, then distinct radial and axial temperatures would have unambiguous meaning. However, as our BO+WKB treatment appreciates coupling between the radial and axial degrees of freedom inherent in the potential, clarification of these quantities is necessary. Here we let $T_\rho$ and $T_z$ determine the distribution according to
\begin{equation}
p_{n_\rho ln_z}=Ne^{-\left(E_{n_\rho ln_z}-E_{00n_z}\right)/k_BT_\rho}e^{-E_{00n_z}/k_BT_z},
\label{Eq:distrotwotemp}
\end{equation}
together with the restriction $E_{n_\rho ln_z}<0$. Here $N$ is a normalization factor, ensuring that the $p_{n_\rho ln_z}$ sum to unity. The distribution given by Eq.~(\ref{Eq:distrotwotemp}) has three noteworthy characteristics. i) The distribution over states of a given $n_z$ follows a Boltzmann distribution with temperature $T_\rho$. This is consistent with the assumption of the previous section. ii) In the limit that $T_\rho$ and $T_z$ are equal, the distribution over all states follows a Boltzmann distribution with temperature $T$, where $T=T_\rho=T_z$ is the singular temperature. This recovers the scenario assumed in the second half of Section~\ref{Sec:BO+WKBlightshifts}. iii) In the limit that the radial and axial degrees of freedom are decoupled, $T_\rho$ and $T_z$ connect with the expected meanings of radial and axial temperature. This is the case, for example, if we substitute the harmonic oscillator potential in place of the full potential, $U\left(\rho,z\right)\rightarrow U^\mathrm{HO}\left(\rho,z\right)$ (and, consequently, $E_{n_\rho ln_z}\rightarrow E^\mathrm{HO}_{n_\rho ln_z}$). While the distribution of Eq.~(\ref{Eq:distrotwotemp}) is contrived to some extent, it nevertheless is expected to provide a reasonable approximation to the experimental conditions in Brown {\it et al.} For the BO+WKB model, Eq.~(\ref{Eq:Xexpression}) for $X$ is modified to
\begin{equation}
X=
\frac{\displaystyle\sum_{n_z}Q_{n_z}\int_{0}^{R_{n_z}(0)}x_{n_z}(\rho)\,\rho\left(e^{-U_{n_z}(\rho)/k_BT_\rho}-1\right)d\rho}
{\displaystyle\sum_{n_z}Q_{n_z}\int_{0}^{R_{n_z}(0)}\rho\left(e^{-U_{n_z}(\rho)/k_BT_\rho}-1\right)d\rho},
\label{Eq:X2temp}
\end{equation}
where
\begin{equation*}
Q_{n_z}=e^{U_{n_z}(0)\left(1/k_BT_\rho-1/k_BT_z\right)}.
\end{equation*}
Analogous expressions hold for $Y$ and $Z$.

In this section, we also consider the Brown {\it et al.}\ model for a distribution characterized by distinct radial and axial temperatures. The corresponding expression for $P_{n_z}$ can be found in Appendix~\ref{Sec:Brownetafac}, which supplements the expressions for $X_{n_z}$, $Y_{n_z}$, and $Z_{n_z}$ given in the previous section. The Ushijima {\it et al.}\ models are not considered here, as this would require making assumptions about how $P_{n_z}$ should be determined for those models in the present case.

\begin{figure*}[tb]
\subfloat{\includegraphics[width=160pt]{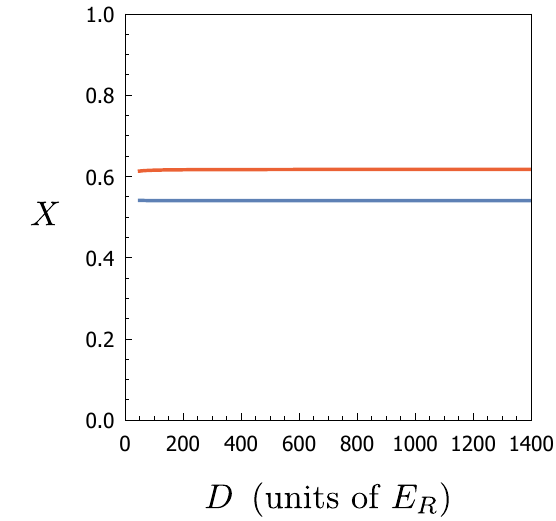}}%
\hspace{15pt}%
\subfloat{\includegraphics[width=160pt]{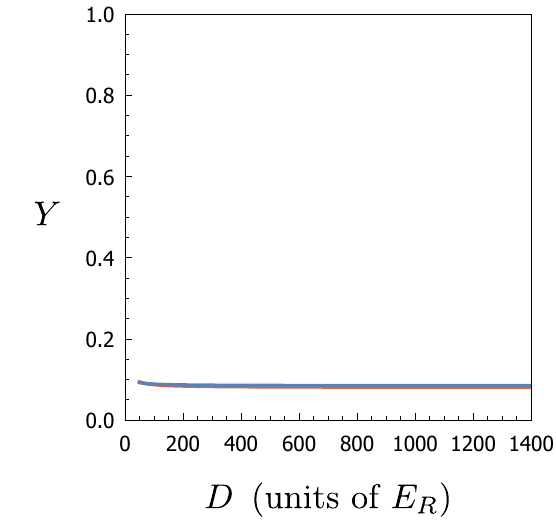}}%
\hspace{15pt}%
\subfloat{\includegraphics[width=160pt]{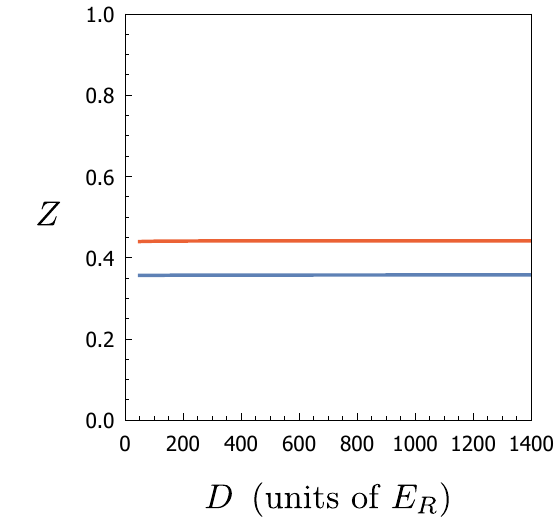}}%
\caption{Factors $X$, $Y$, $Z$ assuming a radial temperature $k_BT_\rho=0.6D$ and an axial temperature $k_BT_z=0.3D$. Results from the BO+WKB model (blue) and the Brown {\it et al.}\ model (red) are both displayed. The lowest depth plotted in each case is $D=50E_R$.}
\label{Fig:XYZforYb}
\end{figure*}

In Brown {\it et al.}, sideband fitting suggests an approximate relationship
$k_BT_z\approx0.3D$ over the range of depths considered. This supplements the approximate relationship $k_BT_\rho\approx0.6D$ noted in the previous section. Figure~\ref{Fig:XYZforYb} presents the resulting $X$, $Y$, and $Z$ versus depth. Consistent with the observation for $X_{n_z}$, $Y_{n_z}$, and $Z_{n_z}$ in the previous section, the BO+WKB model and the Brown~{\it et al.}\ model exhibit similar functional behavior for each of the factors (in the case of $Y$, the curves are largely indistinguishable on the displayed scale). In particular, for the assumed conditions, both models yield factors $X$, $Y$, and $Z$ that are approximately constant over a large range of depths. In Brown {\it et al.}, this motivated reparametrization of Eq.~(\ref{Eq:dEXYZ}) into the simple form
\begin{equation}
\Delta E=a\,D+b\,D^2,
\label{Eq:dEsimpleab}
\end{equation}
with coefficients $a$ and $b$ being independent of depth. We temporarily suppose that $X$, $Y$, and $Z$ are precisely constant. In this case, the reparametrization is exact, with $a$ and $b$ given by
\begin{gather*}
a=-\frac{\Delta\alpha_{E1}}{\alpha_{E1}}X-\frac{\Delta\alpha_{M1+E2}}{\alpha_{E1}}Y,
\\
b=-\frac{\Delta\beta}{\alpha_{E1}^2}Z.
\end{gather*}
Due to the presence of $\Delta\alpha_{E1}$, the coefficient $a$ is dependent on the lattice frequency. As with $\Delta\alpha_{E1}$, it can be decomposed in terms of a slope and a zero-crossing with respect to the lattice frequency. Given this decomposition, $\Delta\alpha_{M1+E2}$ merely manifests itself as a displacement of the zero-crossing away from the magic frequency. Meanwhile, the coefficient $b$ is independent of the lattice frequency.

In Brown {\it et al.}, Eq.~(\ref{Eq:dEsimpleab}) was used to characterize the lattice light shifts in a Yb optical lattice clock. Individual measurements involved interleaving between a test depth and a reference depth, with the lattice frequency held fixed, while recording the difference in the clock frequency. A common reference depth ($\approx\!180E_R$) was used for all measurements, while several test depths and lattice frequencies were explored. The coefficients $a$ (slope and zero-crossing) and $b$ of Eq.~(\ref{Eq:dEsimpleab}) were determined by a least-squares fitting with the experimental data. For a given operational depth and lattice frequency, the lattice light shift is then evaluated from Eq.~(\ref{Eq:dEsimpleab}) using the experimentally-determined coefficients. In applying this procedure, however, it's necessary to appreciate that $X$, $Y$, and $Z$ are not precisely constant over the range of depths considered. Consequently, the parametrization of Eq.~(\ref{Eq:dEsimpleab}) is inexact, and this may lead to error in the evaluated lattice light shift.

To estimate the size of this error, we generate mock data from the BO+WKB model. The mock data is intended to imitate the experimentally-obtained data. It encompasses a similar range of test depths and lattice frequencies, while using the same reference depth. With this mock data set, we apply the above procedure (i.e., determining the coefficients of Eq.~(\ref{Eq:dEsimpleab}) through least squares fitting) to evaluate the lattice light shift for the operational depth and lattice frequency. Separately, the lattice light shift for this depth and lattice frequency is evaluated directly with the BO+WKB model. The two results are compared, with the difference being indicative of the error associated with the reparametrization~\cite{reparamnote}. To perform this assessment, we must assume values for the atomic parameters of Yb. The atomic parameters can be estimated from the experimentally-obtained coefficients $a$ and $b$, together with the (nearly constant) theoretical values of $X$, $Y$, and $Z$. However, since we don't possess an independent measure of the magic frequency to compare with the experimental zero-crossing of $a$, we cannot extract the parameter $\Delta\alpha_{M1+E2}/\alpha_{E1}$ in this way. Instead, we take the theoretical value of $\Delta\alpha_{M1+E2}/\alpha_{E1}$ given in Brown~{\it et al.} For an operational depth of $50E_R$ and an operational lattice frequency chosen to give a stable lattice light shift with respect to fluctuations in the depth (corresponding to the ``operational magic wavelength'' discussed in Brown~{\it et al.}), we find the resulting error to be below $10^{-19}$.

We note that the reparametrization above is motivated entirely by the functional form of $X$, $Y$, and $Z$ versus depth (namely, being approximately constant) and not by the precise values of these factors. This relaxes the need to be quantitatively precise about the distribution over motional states. For example, the near-constancy of $X$, $Y$, and $Z$ observed in Fig.~\ref{Fig:XYZforYb} is more a consequence of the proportionality relation $T_\rho,T_z\propto D$ than it is of the specific proportionality factors connecting the temperatures to depth. Thus, so long as important qualitative features of the distribution and its dependence on depth are captured, reparametrized forms of the lattice light shift can have broad applicability, while also relaxing the need for precise knowledge of the atomic parameters. On the flip side, this makes it challenging to compare or translate parameters between experiments. For example, the coefficients $a$ (slope and zero crossing) and $b$ in Eq.~(\ref{Eq:dEsimpleab}) are not universal. Despite the extra leeway that can be afforded by reparametrization, it is still important to account for uncertainty in the distribution. For example, in the present case we consider a ``spread'' of possible distributions by introducing non-linear (offset and quadratic) terms into the relationship for $T_\rho$ and $T_z$ versus $D$. Different coefficients are explored for these relationships, with the spread guided by the sideband fitting. Following the procedure outlined in the previous paragraph, we find that reparametrization error at the low-$10^{-18}$ level can result in some cases when using Eq.~(\ref{Eq:dEsimpleab}). If higher accuracy is sought, the error can be reduced by introducing additional terms into Eq.~(\ref{Eq:dEsimpleab}), which comes at the cost of an increased number of fitting parameters. Further discussion (in the context of the Brown {\it et al.}\ model) can be found in the main text and supplementary material of Ref.~\cite{BroPhiBel17}.

Following the discussion above, it's clear that if less-sophisticated models are to be used to motivate reparametrization of the lattice light shift and to assess the corresponding error, an emphasis should be put on accurate representation of the functional form of $X$, $Y$, and $Z$ versus $D$ rather than the absolute accuracy of these factors. We find that as we vary the distribution, the functional behavior exhibited by the Brown {\it et al.}\ model closely matches that of the BO+WKB model (largely differing by a scale factor for $X$, $Y$, and $Z$). Thus, while the Brown {\it et al.}\ model isn't expected to support high-accuracy in absolute terms, it is nevertheless capable of supporting high-accuracy for the purposes of reparametrizing the lattice light shift. Here we have essentially used the more-sophisticated BO+WKB model to validate the Brown {\it et al.}\ model in a specific context.

Before concluding this section, we note that there also exists a discrepancy for the atomic parameter $\Delta\alpha_{M1+E2}/\alpha_{E1}$ in Yb. Nemitz~{\it et al.}~\cite{NemJorYan19} recently reported a value that differs by a few $\sigma$, where $\sigma$ represents combined standard deviation, from the theoretical value given in Brown~{\it et al.} To extract this parameter from experimental measurements, Nemitz~{\it et al.}\ employed a model that is closely related to the Ushijima~{\it et al.}\ model, but also incorporates aspects of the sideband fitting model of Ref.~\cite{BlaThoCam09}. Using the value for $\Delta\alpha_{M1+E2}/\alpha_{E1}$ given by Nemitz~{\it et al.}, which is larger in absolute value, we note that the reparametrization errors tend to get worse. More theoretical or experimental investigation will be required to resolve this issue.

\section{Conclusion}
\label{Sec:Conclusion}

Here we have developed a non-perturbative model to describe the motional energy levels in a one-dimensional optical lattice, accounting for both axial and radial confinement relative to the lattice axis. The model combines established tools of quantum mechanics: the Born-Oppenheimer approximation and the WKB approximation. We extend our model to describe lattice light shifts in optical lattice clocks. To evaluate the lattice light shift, a distribution over motional states needs to be specified for the atomic ensemble. We consider distributions that are expected to approximate experimental conditions, with corresponding expressions being presented. Comparisons are provided between our BO+WKB model and other lattice light shift models from the literature. We believe our BO+WKB model can be a valuable tool for evaluating lattice light shifts or for assessing the validity of simpler models under certain conditions.

Like other lattice light shift models in the literature, our BO+WKB model is built upon some idealizations. As discussed in the text, tunneling between lattice sites is neglected. While it is noted that gravity can help suppress tunneling in practice, we do not formally account for gravity. A component of gravity perpendicular to the lattice axis can introduce gravitational sag, with the atoms effectively being pulled away from the lattice axis. Also, some degree of imperfection will inevitably be present in the optical lattice. This may include a misbalance of intensity or a misalignment of the axes or polarization vectors of the counter-propagating waves that form the optical lattice. We also neglect site-to-site variations in the optical lattice due, for example, to focusing. In principle, sideband (non-carrier) transitions may lead to line-pulling. While transitions connecting motional states with different quantum number $n_\rho$ are nominally forbidden when the clock laser is aligned with the lattice axis, this is not strictly true due to motional coupling between the radial and axial degrees of freedom. In optical lattice clocks seeking high accuracy, effects such as these warrant additional consideration.

Finally, while our BO+WKB model provides a means to evaluate lattice light shifts when the motional state distribution is known, it does not address how this distribution should be determined. Techniques such as sideband fitting~\cite{BlaThoCam09,NemJorYan19} and time-of-flight thermometry~\cite{UshTakKat18} have proven useful for this purpose, though the integrity of these methods has not been well-assessed. For example, the sideband fitting protocols are based on a perturbative approach starting from the harmonic oscillator approximation, for which we have expressed skepticism in this work. In the previous section, a strategy for characterizing lattice light shifts is outlined that relies on qualitative aspects of the sideband fitting, but not on precise results. Nevertheless, with the goal of reducing lattice light shift uncertainty further, there is strong motivation for developing methods for accurately characterizing the motional state distribution in the optical lattice. Alternatively, optical lattice clocks based on three-dimensional optical lattices~\cite{AkaTakKat08,CamHutMar17} can circumvent this issue altogether, allowing the atoms to be unambiguously prepared in the ground motional state. While three dimensional optical lattice clocks certainly come with their own set of metrological challenges, it is an outstanding question whether these challenges prove to be less of an impedance than the need to characterize the motional state distribution in one dimensional lattices.

\acknowledgments

We thank B.\ Patla and E.\ Clements for their careful reading of the manuscript. This work was supported by NIST, DARPA, and the NASA Fundamental Physics Program.

\appendix

\section{Harmonic oscillator energies and wave functions}
\label{Sec:HOwf}

The harmonic oscillator potential, Eq.~(\ref{Eq:HOpot}), can be written in the form
\begin{equation*}
U^\mathrm{HO}(\rho,z)=-D+\frac{1}{2}m\Omega_\rho^2\rho^2+\frac{1}{2}m\Omega_z^2z^2.
\end{equation*}
The energies and wave functions for this potential are
\begin{gather*}
E^\mathrm{HO}_{n_\rho l n_z}=
-D
+\hbar\Omega_\rho\left(2n_\rho+\left|l\right|+1\right)
+\hbar\Omega_z\left(n_z+\frac{1}{2}\right),
\\
\begin{aligned}
\Psi^\mathrm{HO}_{n_\rho ln_z}(\rho,\varphi,z)=
{}&
\frac{b_\rho\sqrt{b_z}}{\pi^{3/4}}\sqrt{\frac{n_\rho!}{(n_\rho+|l|)!n_z!2^{n_z}}}\left(b_\rho\rho\right)^{|l|}
\\&\times
\exp\left(-\frac{b_\rho^2\rho^2}{2}\right)
L_{n_\rho}^{|l|}\left(b_\rho^2\rho^2\right)e^{il\varphi}
\\&\times
\exp\left(-\frac{b_z^2z^2}{2}\right)
H_{n_z}\left(b_zz\right),
\end{aligned}
\end{gather*}
where $b_\rho\equiv\sqrt{m\Omega_\rho/\hbar}$ and $b_z\equiv\sqrt{m\Omega_z/\hbar}$. The functions $L_n^k(x)$ and $H_n(x)$ are associated Laguerre polynomials and Hermite polynomials, respectively. These wave functions satisfy
\begin{gather*}
\int_{-\infty}^{+\infty}\int_{0}^{2\pi}\int_0^\infty\Psi^\mathrm{HO*}_{n_\rho l n_z}(\rho,\varphi,z)\Psi^\mathrm{HO}_{n_\rho^\prime l^\prime n_z^\prime}(\rho,\varphi,z)\rho\,d\rho\,d\varphi\,dz
\\
=\delta_{n_\rho n_\rho^\prime}\delta_{ll^\prime}\delta_{n_zn_z^\prime}.
\end{gather*}

The perturbative approach described in Sec.~\ref{Sec:HOapprox} requires evaluation of the matrix elements $\bigl\langle\Psi^\mathrm{HO}_{n_\rho l n_z}\bigl|\rho^pz^q\bigr|\Psi^\mathrm{HO}_{n_\rho^\prime l^\prime n_z^\prime}\bigr\rangle$, where $p$ and $q$ are even, positive integers. Here $\bigl\langle\Psi^\mathrm{HO}_{n_\rho l n_z}\bigl|\rho^pz^q\bigr|\Psi^\mathrm{HO}_{n_\rho^\prime l^\prime n_z^\prime}\bigr\rangle$ denotes the integral above, but with the inclusion of $\rho^pz^q$ in the integrand. The matrix elements of $\rho^2$ and $z^2$ read
\begin{align*}
\bigl\langle\Psi^\mathrm{HO}_{n_\rho l n_z}\bigl|\rho^2\bigr|\Psi^\mathrm{HO}_{n_\rho^\prime l^\prime n_z^\prime}\bigr\rangle={}&
\delta_{ll^\prime}\delta_{n_z n_z^\prime}\frac{\hbar}{m\Omega_\rho}
\\&\times
\biggl[\delta_{n_\rho n_\rho^\prime}\left(2n_\rho+\left|l\right|+1\right)
\\&
-\delta_{n_\rho,n_\rho^\prime+1}\sqrt{n_\rho\left(n_\rho+\left|l\right|\right)}
\\&
-\delta_{n_\rho+1,n_\rho^\prime}\sqrt{n_\rho^\prime\left(n_\rho^\prime+\left|l\right|\right)}\biggr],
\\
\bigl\langle\Psi^\mathrm{HO}_{n_\rho l n_z}\bigl|z^2\bigr|\Psi^\mathrm{HO}_{n_\rho^\prime l^\prime n_z^\prime}\bigr\rangle
={}&
\delta_{n_\rho n_\rho^\prime}\delta_{ll^\prime}\frac{\hbar}{2m\Omega_z}
\Bigl[\delta_{n_zn_z^\prime}\left(2n_z+1\right)
\\&
+\delta_{n_z,n_z^\prime+2}\sqrt{n_z\left(n_z-1\right)}
\\&
+\delta_{n_z+2,n_z^\prime}\sqrt{n_z^\prime\left(n_z^\prime-1\right)}
\Bigr].
\end{align*}
The matrix element for $z^2$ is readily obtained with raising and lowering operators for a one-dimensional harmonic oscillator. The matrix element for $\rho^2$ was derived from orthogonality and recurrence relations of the associated Laguerre polynomials. Exploiting completeness of the harmonic oscillator states, expressions for the matrix elements $\bigl\langle\Psi^\mathrm{HO}_{n_\rho l n_z}\bigl|\rho^pz^q\bigr|\Psi^\mathrm{HO}_{n_\rho^\prime l^\prime n_z^\prime}\bigr\rangle$ with $p+q>2$ can further be obtained from these expressions.

\section{Density of states for the harmonic oscillator potential}
\label{Sec:HODOS}

For $\gamma\gg1$, there are $\approx\!\gamma^2/2$ combinations of $n_\rho$ and $l$ that satisfy $2n_\rho+|l|+1<\gamma$. We introduce the function
\begin{gather*}
\gamma_{n_z}(E)=\frac{\displaystyle E+D-2\sqrt{DE_R}\left(n_z+\frac{1}{2}\right)}{2\sqrt{DE_R}(\kappa/k)},
\end{gather*}
which satisfies $\gamma_{n_z}(E^\mathrm{HO}_{n_\rho ln_z})=2n_\rho+|l|+1$. We subsequently introduce
\begin{gather*}
\mathcal{N}_{n_z}(E)=\frac{1}{2}\left[\gamma_{n_z}(E)\right]^2,
\end{gather*}
which represents the total number of harmonic oscillator states for a given $n_z$ that have an energy below $E$. The density of states is then taken as $G^\mathrm{HO}_{n_z}(E)=\mathcal{N}_{n_z}^\prime(E)$, which results in Eq.~(\ref{Eq:HODOS}) of the main text.

\section{Orthogonality and normalization of the BO wave functions}
\label{Sec:BOnorm}

Eigenfunctions of the axial BO equation, Eq.~(\ref{Eq:axialBO}), are orthogonal, and we take them to be normalized to unity. That is, the eigenfunctions satisfy
\begin{equation*}
\int_{-\pi/2k}^{+\pi/2k} \mathcal{Z}^*_{n_z}(\rho,z)\mathcal{Z}_{n_z^\prime}(\rho,z)dz=\delta_{n_zn_z^\prime}. 
\end{equation*}
The integration limits are taken as $\pm\pi/2k$ due to the assumption of infinite potential barriers for $|z|>\pi/2k$ (see Sec.~\ref{Sec:BO}). It is noted in Sec.~\ref{Sec:HOrevisited} that the BO approach is exact for the harmonic oscillator potential, Eq.~(\ref{Eq:HOpot}). In that case, it is understood that the integration limits extend to $\pm\infty$. For a given $n_z$ and $l$, eigenfunctions of the radial BO equation, Eq.~(\ref{Eq:radialBO}), are orthogonal, and we can likewise take them to be normalized to unity,
\begin{equation*}
\int_0^\infty \mathcal{R}^*_{n_\rho l n_z}(\rho)\mathcal{R}_{n_\rho^\prime l n_z}(\rho) d\rho=\delta_{n_\rho n_\rho^\prime}. 
\end{equation*}
The BO wave functions, Eq.~(\ref{Eq:BOwf}), subsequently satisfy 
\begin{gather*}
\int_{-\pi/2k}^{+\pi/2k}\int_{0}^{2\pi}\!\int_0^\infty\!\Psi^*_{n_\rho l n_z}(\rho,\varphi,z)\Psi_{n_\rho^\prime l^\prime n_z^\prime}(\rho,\varphi,z)\rho\,d\rho\,d\varphi\,dz
\\
=\delta_{n_\rho n_\rho^\prime}\delta_{ll^\prime}\delta_{n_zn_z^\prime}.
\end{gather*}

\section{Solutions to the axial BO equation}
\label{Sec:Zsol}

The axial BO equation, Eq.~(\ref{Eq:axialBO}), taken together with the potential, Eq.~(\ref{Eq:potential}), amounts to Mathieu's differential equation. Insisting that the eigenfunctions vanish at $|z|=\pi/2k$, the solutions are
\begin{gather}
\mathcal{Z}_{n_z}(\rho,z)=
\sqrt{\frac{2k}{\pi}}\times
\left\{\begin{array}{c}
{\displaystyle ce_{n_z+1}\left(kz,-\frac{\mathcal{D}(\rho)}{4}\right)}
\\
{\displaystyle se_{n_z+1}\left(kz,-\frac{\mathcal{D}(\rho)}{4}\right)}
\end{array}\right.,
\label{Eq:Z1}
\\
U_{n_z}(\rho)=
E_R\times
\left\{\begin{array}{c}
{\displaystyle a_{n_z+1}\left(-\frac{\mathcal{D}(\rho)}{4}\right)-\frac{\mathcal{D}(\rho)}{2}}
\\
{\displaystyle b_{n_z+1}\left(-\frac{\mathcal{D}(\rho)}{4}\right)-\frac{\mathcal{D}(\rho)}{2}}
\end{array}\right.,
\nonumber
\end{gather}
where the top (bottom) lines apply for even (odd) $n_z$ and where we introduced $\mathcal{D}(\rho)=(D/E_R)e^{-\kappa^2\rho^2}$ for brevity. Here $ce_r(z,q)$ and $se_r(z,q)$ are even and odd Mathieu functions, respectively, with $a_r(q)$ and $b_r(q)$ being their respective characteristic values~\cite{Bra72}.

We can exploit properties of the Mathieu functions and their characteristic values to instead express the eigenfunctions and eigenvalues as
\begin{gather}
\mathcal{Z}_{n_z}(\rho,z)=
\sqrt{\frac{2k}{\pi}}
se_{n_z+1}\left(kz+\frac{\pi}{2},\frac{\mathcal{D}(\rho)}{4}\right),
\label{Eq:Z2}
\\
U_{n_z}(\rho)=
E_R\left[b_{n_z+1}\left(\frac{\mathcal{D}(\rho)}{4}\right)-\frac{\mathcal{D}(\rho)}{2}\right],
\nonumber
\end{gather}
which is applicable for both even and odd $n_z$. In going from expression (\ref{Eq:Z1}) to expression (\ref{Eq:Z2}) for the eigenfunctions, we have dropped an $n_z$-dependent sign.

\section{Integrations with respect to $l$}
\label{Sec:lint}

Here we consider the integral
\begin{equation*}
\mathcal{I}=\int_\mathbb{R}\frac{1}{\displaystyle\sqrt{E-U_{n_z}(\rho)-\frac{\hbar^2}{2m}\frac{l^2}{\rho^2}}}dl.
\end{equation*}
For this integral to return a non-zero value, we must have $E>U_{n_z}(\rho)$. When this criterion is satisfied, we may introduce a variable $u$ given by
\begin{equation*}
u=\frac{l}{\rho\displaystyle\sqrt{\frac{2m}{\hbar^2}\left[E-U_{n_z}(\rho)\right]}}.
\end{equation*}
In terms of $u$, the integral becomes
\begin{equation*}
\mathcal{I}=\rho\sqrt{\frac{2m}{\hbar^2}}\int_{-1}^{+1}\frac{1}{\sqrt{1-u^2}}du,
\end{equation*}
where the integration limits are easily deduced. The definite integral here equates to $\pi$. It follows that
\begin{equation*}
\mathcal{I}=\pi\sqrt{\frac{2m}{\hbar^2}}\rho,
\end{equation*}
if $E>U_{n_z}(\rho)$ and $\mathcal{I}=0$ otherwise.

\section{Analytical expression for the numerator of Eq.~(\ref{Eq:Xexpression})}
\label{Sec:ananum}

Here we derive an analytical expression for the integral
\begin{equation*}
\mathcal{I}=\frac{2\kappa^2D}{k_BT}\int_{0}^{R_{n_z}(0)}x_{n_z}(\rho)\,\rho\left[\exp\left(-\frac{U_{n_z}(\rho)}{k_BT}\right)-1\right]d\rho,
\end{equation*}
which appears (without prefactor) in the numerator of Eqs.~(\ref{Eq:Xexpression}), (\ref{Eq:Xn}), and (\ref{Eq:X2temp}). As a first step, we note the equality
\begin{equation*}
U_{n_z}^\prime(\rho)
=\int_{-\pi/2k}^{+\pi/2k}\left|\mathcal{Z}_{n_z}(\rho,z)\right|^2\frac{\partial U(\rho,z)}{\partial\rho}dz,
\end{equation*}
which follows from the Hellmann-Feynman theorem. Given the form of $U(\rho,z)$ and the form of $x_{n_z}(\rho)$, this implies
\begin{equation*}
x_{n_z}(\rho)=\frac{U_{n_z}^\prime(\rho)}{2\kappa^2D\rho}.
\end{equation*}
Thus, we can write
\begin{equation*}
\mathcal{I}=\int_{0}^{R_{n_z}(0)}\left[\exp\left(-\frac{U_{n_z}(\rho)}{k_BT}\right)-1\right]\frac{U_{n_z}^\prime(\rho)}{k_BT}d\rho.
\end{equation*}
Making the substitution $u=U_{n_z}(\rho)/k_BT$, this integral becomes
\begin{equation*}
\mathcal{I}=\int_{u_0}^{0}\left(e^{-u}-1\right)du,
\end{equation*}
where $u_0=U_{n_z}(0)/k_BT$. This integral is readily evaluated, giving
\begin{equation*}
\mathcal{I}=e^{-u_0}+u_0-1.
\end{equation*}

\section{Supplemental information for the Brown {\it et al.}\ model}
\label{Sec:Brownetafac}

Here we provide supplemental information for the Brown {\it et al.}\ model, as not all details were explicit in that work. As noted in Sec.~\ref{Sec:comparemodels}, an expression is provided in Brown {\it et al.}\ for the clock shift of a specific motional state. This expression is in terms of quantum numbers $n_r$ and $n_z$, where $n_r\equiv 2n_\rho+\left|l\right|$ (in Brown {\it et al.}, $n_r$ is written as $n_\rho$, but we have a different meaning for $n_\rho$ in this work). To completely define the Brown {\it et al.}\ model, we must further specify the energies used to determine the populations (e.g., Eqs.~(\ref{Eq:XnzTrho}) and (\ref{Eq:distrotwotemp})). For this purpose, the following energies are used
\begin{equation}
\begin{aligned}
E^\mathrm{Brown}_{n_rn_z}={}&
-D+2\sqrt{DE_R}\left(\kappa/k\right)\left(n_r+1\right)
\\&
-\frac{2}{3}E_R\left(\kappa/k\right)^2\left(n_r^2+2n_r+\frac{3}{2}\right)
\\&
+2\sqrt{DE_R}\left(n_z+\frac{1}{2}\right)
\\&
-\frac{1}{2}E_R\left(n_z^2+n_z+\frac{1}{2}\right).
\end{aligned}
\label{Eq:EBrown}
\end{equation}
These energies are derived using the perturbative approach described in Sec.~\ref{Sec:HOapprox}. Specifically, terms proportional to $\rho^4$ and $z^4$ in the series expansion of the potential are evaluated at first order in perturbation theory. For the radial anharmonic correction, an average is further taken over all states of a given $n_r$ to arrive at the above expression (this step was also performed in arriving at the clock shift expression in Brown~{\it et al.}). No cross-dimensional corrections are included in Eq.~(\ref{Eq:EBrown}). Given that the radial and axial degrees of freedom remain decoupled in this picture, $T_\rho$ and $T_z$ of the main text connect with the conventional meaning of radial and axial temperature. 

In the clock shift expression of Brown {\it et al.}, $n_r$ appears as $n_r^p$, with the exponent $p$ equal to one or two. In deducing expressions for $X_{n_z}$, $Y_{n_z}$, and $Z_{n_z}$, we must evaluate the average value of $n_r^p$ for a given $n_z$. We write this as $\left\langle n_r^p\right\rangle_{n_z}$, and it is taken as
\begin{equation}
\left\langle n_r^p\right\rangle_{n_z}=\frac{\displaystyle\sum_{n_r}n_r^p(n_r+1)e^{-E^\mathrm{Brown}_{n_rn_z}/k_BT_\rho}}{\displaystyle\sum_{n_r}(n_r+1)e^{-E^\mathrm{Brown}_{n_rn_z}/k_BT_\rho}}.
\label{Eq:nravg}
\end{equation}
The factor $(n_r+1)$ appearing in the numerator and denominator on the right-hand-side is a degeneracy factor accounting for all states of a given $n_r$ (i.e., all combinations of $n_\rho$ and $l$ compatible with the given $n_r$). While $n_z$ does appear explicitly on the right-hand-side, we note that this does not contribute to an $n_z$-dependence, since the axial contribution to $E^\mathrm{Brown}_{n_rn_z}$ simply cancels out of the numerator and denominator. Rather, the $n_z$-dependence arises from the implicit restriction $E^\mathrm{Brown}_{n_rn_z}<0$ on the summations, limiting the included states to those representative of trapped motional states. Starting from the ground state, $n_r=n_z=0$, the energies increase monotonically with $n_r$ and $n_z$ until they exceed zero. According to Eq.~(\ref{Eq:EBrown}), as $n_r$ and $n_z$ are further increased, the energies return to the negative-energy region and continue on to negative infinity. These additional negative-energy states do not correspond to physical states of the potential, and it is understood that they are also excluded from the summations.

Given $\kappa/k\ll1$, we replace the summations in Eq.~(\ref{Eq:nravg}) with integrals. In doing so, we arrive at the expression
\begin{equation}
\left\langle n_r^p\right\rangle_{n_z}=
\frac{\left(p+1\right)!}{2^p}\left(\frac{\kappa}{k}\right)^{-p}\left(\frac{k_BT_\rho}{D}\right)^{p}\left(\frac{D}{E_R}\right)^{p/2}
\eta^{(p)}_{n_z},
\label{Eq:meannr}
\end{equation}
where $\eta^{(p)}_{n_z}=\Lambda^{(p+1)}_{n_z}/\Lambda^{(1)}_{n_z}$ is the factor appearing in Sec.~\ref{Sec:comparemodels}. Here $\Lambda^{(q)}_{n_z}$ is given by
\begin{equation*}
\Lambda^{(q)}_{n_z}=
\frac{1}{q!}\int_0^{\chi_{n_z}}\chi^{q}\exp\left[-\chi+\frac{1}{6}\left(\frac{k_BT_\rho}{D}\right)\chi^2\right]d\chi,
\end{equation*}
with the upper limit on this integral being
\begin{align*}
\chi_{n_z}={}&
3\left(\frac{k_BT_\rho}{D}\right)^{-1}
\Bigg\{1-\left[\frac{1}{3}+\frac{4}{3}\left(\frac{D}{E_R}\right)^{-1/2}\left(n_z+\frac{1}{2}\right)
\right.\\&\left.
-\frac{1}{3}\left(\frac{D}{E_R}\right)^{-1}\left(n_z^2+n_z+\frac{1}{2}\right)\right]^{1/2}\Bigg\}.
\end{align*}
In the limit $k_BT_\rho/D\rightarrow0$, $\Lambda^{(q)}_{n_z}\rightarrow1$ and $\eta^{(p)}_{n_z}\rightarrow1$. If we had omitted the anharmonic corrections altogether (i.e., $E^\mathrm{Brown}_{n_rn_z}\rightarrow E^\mathrm{HO}_{n_rn_z}$) and allowed $n_r$ to sum freely to infinity in Eq.~(\ref{Eq:nravg}), we would have arrived at Eq.~(\ref{Eq:meannr}) without the factor $\eta^{(p)}_{n_z}$. Thus, $\eta^{(p)}_{n_z}$ may be regarded as a correction factor relative to this case.

For the distribution given by Eq.~(\ref{Eq:distrotwotemp}), an expression for $P_{n_z}$ can be obtained with similar considerations. In particular, we find
\begin{align*}
P_{n_z}={}&\mathcal{N}
\Lambda^{(1)}_{n_z}\exp\Bigg[-2\frac{\sqrt{DE_R}}{k_BT_z}\left(n_z+\frac{1}{2}\right)
\\&
-\frac{1}{2}\frac{E_R}{k_BT_z}\left(n_z^2+n_z+\frac{1}{2}\right)\Bigg].
\end{align*}
Here $\mathcal{N}$ is a normalization factor, ensuring that the $P_{n_z}$ sum to unity. Note that even though the radial and axial degrees of freedom are decoupled according to Eq.~(\ref{Eq:EBrown}), the $P_{n_z}$ still depend on the radial temperature $T_\rho$ via the factor $\Lambda^{(1)}_{n_z}$. This dependence is a result of the restriction $E^\mathrm{Brown}_{n_rn_z}<0$. 


\begin{thebibliography}{40}%
\makeatletter
\providecommand \@ifxundefined [1]{%
 \@ifx{#1\undefined}
}%
\providecommand \@ifnum [1]{%
 \ifnum #1\expandafter \@firstoftwo
 \else \expandafter \@secondoftwo
 \fi
}%
\providecommand \@ifx [1]{%
 \ifx #1\expandafter \@firstoftwo
 \else \expandafter \@secondoftwo
 \fi
}%
\providecommand \natexlab [1]{#1}%
\providecommand \enquote  [1]{``#1''}%
\providecommand \bibnamefont  [1]{#1}%
\providecommand \bibfnamefont [1]{#1}%
\providecommand \citenamefont [1]{#1}%
\providecommand \href@noop [0]{\@secondoftwo}%
\providecommand \href [0]{\begingroup \@sanitize@url \@href}%
\providecommand \@href[1]{\@@startlink{#1}\@@href}%
\providecommand \@@href[1]{\endgroup#1\@@endlink}%
\providecommand \@sanitize@url [0]{\catcode `\\12\catcode `\$12\catcode
  `\&12\catcode `\#12\catcode `\^12\catcode `\_12\catcode `\%12\relax}%
\providecommand \@@startlink[1]{}%
\providecommand \@@endlink[0]{}%
\providecommand \url  [0]{\begingroup\@sanitize@url \@url }%
\providecommand \@url [1]{\endgroup\@href {#1}{\urlprefix }}%
\providecommand \urlprefix  [0]{URL }%
\providecommand \Eprint [0]{\href }%
\providecommand \doibase [0]{http://dx.doi.org/}%
\providecommand \selectlanguage [0]{\@gobble}%
\providecommand \bibinfo  [0]{\@secondoftwo}%
\providecommand \bibfield  [0]{\@secondoftwo}%
\providecommand \translation [1]{[#1]}%
\providecommand \BibitemOpen [0]{}%
\providecommand \bibitemStop [0]{}%
\providecommand \bibitemNoStop [0]{.\EOS\space}%
\providecommand \EOS [0]{\spacefactor3000\relax}%
\providecommand \BibitemShut  [1]{\csname bibitem#1\endcsname}%
\let\auto@bib@innerbib\@empty
\bibitem [{\citenamefont {Katori}(2002)}]{Kat02}%
  \BibitemOpen
  \bibfield  {author} {\bibinfo {author} {\bibfnamefont {H.}~\bibnamefont
  {Katori}},\ }\href {\doibase 10.1142/9789812777713_0036} {\emph {\bibinfo
  {title} {\emph{in} Proceedings of the 6th Symposium Frequency Standards and
  Metrology}}},\ edited by\ \bibinfo {editor} {\bibfnamefont {P.}~\bibnamefont
  {Gill}}\ (\bibinfo  {publisher} {World Scientific, Singapore},\ \bibinfo
  {year} {2002})\ p.\ \bibinfo {pages} {323}\BibitemShut {NoStop}%
\bibitem [{\citenamefont {Katori}\ \emph {et~al.}(2003)\citenamefont {Katori},
  \citenamefont {Takamoto}, \citenamefont {Pal'chikov},\ and\ \citenamefont
  {Ovsiannikov}}]{KatTakPal03}%
  \BibitemOpen
  \bibfield  {author} {\bibinfo {author} {\bibfnamefont {H.}~\bibnamefont
  {Katori}}, \bibinfo {author} {\bibfnamefont {M.}~\bibnamefont {Takamoto}},
  \bibinfo {author} {\bibfnamefont {V.~G.}\ \bibnamefont {Pal'chikov}}, \ and\
  \bibinfo {author} {\bibfnamefont {V.~D.}\ \bibnamefont {Ovsiannikov}},\
  }\href {\doibase 10.1103/PhysRevLett.91.173005} {\bibfield  {journal}
  {\bibinfo  {journal} {Phys. Rev. Lett.}\ }\textbf {\bibinfo {volume} {91}},\
  \bibinfo {pages} {173005} (\bibinfo {year} {2003})}\BibitemShut {NoStop}%
\bibitem [{\citenamefont {Takamoto}\ and\ \citenamefont
  {Katori}(2003)}]{TakKat03}%
  \BibitemOpen
  \bibfield  {author} {\bibinfo {author} {\bibfnamefont {M.}~\bibnamefont
  {Takamoto}}\ and\ \bibinfo {author} {\bibfnamefont {H.}~\bibnamefont
  {Katori}},\ }\href {\doibase 10.1103/PhysRevLett.91.223001} {\bibfield
  {journal} {\bibinfo  {journal} {Phys. Rev. Lett.}\ }\textbf {\bibinfo
  {volume} {91}},\ \bibinfo {pages} {223001} (\bibinfo {year}
  {2003})}\BibitemShut {NoStop}%
\bibitem [{\citenamefont {Barber}\ \emph {et~al.}(2006)\citenamefont {Barber},
  \citenamefont {Hoyt}, \citenamefont {Oates}, \citenamefont {Hollberg},
  \citenamefont {Taichenachev},\ and\ \citenamefont {Yudin}}]{BarHoyOat06}%
  \BibitemOpen
  \bibfield  {author} {\bibinfo {author} {\bibfnamefont {Z.~W.}\ \bibnamefont
  {Barber}}, \bibinfo {author} {\bibfnamefont {C.~W.}\ \bibnamefont {Hoyt}},
  \bibinfo {author} {\bibfnamefont {C.~W.}\ \bibnamefont {Oates}}, \bibinfo
  {author} {\bibfnamefont {L.}~\bibnamefont {Hollberg}}, \bibinfo {author}
  {\bibfnamefont {A.~V.}\ \bibnamefont {Taichenachev}}, \ and\ \bibinfo
  {author} {\bibfnamefont {V.~I.}\ \bibnamefont {Yudin}},\ }\href {\doibase
  10.1103/PhysRevLett.96.083002} {\bibfield  {journal} {\bibinfo  {journal}
  {Phys. Rev. Lett.}\ }\textbf {\bibinfo {volume} {96}},\ \bibinfo {pages}
  {083002} (\bibinfo {year} {2006})}\BibitemShut {NoStop}%
\bibitem [{\citenamefont {Yi}\ \emph {et~al.}(2011)\citenamefont {Yi},
  \citenamefont {Mejri}, \citenamefont {McFerran}, \citenamefont {Le~Coq},\
  and\ \citenamefont {Bize}}]{YiMejMcF11}%
  \BibitemOpen
  \bibfield  {author} {\bibinfo {author} {\bibfnamefont {L.}~\bibnamefont
  {Yi}}, \bibinfo {author} {\bibfnamefont {S.}~\bibnamefont {Mejri}}, \bibinfo
  {author} {\bibfnamefont {J.~J.}\ \bibnamefont {McFerran}}, \bibinfo {author}
  {\bibfnamefont {Y.}~\bibnamefont {Le~Coq}}, \ and\ \bibinfo {author}
  {\bibfnamefont {S.}~\bibnamefont {Bize}},\ }\href {\doibase
  10.1103/PhysRevLett.106.073005} {\bibfield  {journal} {\bibinfo  {journal}
  {Phys. Rev. Lett.}\ }\textbf {\bibinfo {volume} {106}},\ \bibinfo {pages}
  {073005} (\bibinfo {year} {2011})}\BibitemShut {NoStop}%
\bibitem [{\citenamefont {Kulosa}\ \emph {et~al.}(2015)\citenamefont {Kulosa},
  \citenamefont {Fim}, \citenamefont {Zipfel}, \citenamefont {R\"uhmann},
  \citenamefont {Sauer}, \citenamefont {Jha}, \citenamefont {Gibble},
  \citenamefont {Ertmer}, \citenamefont {Rasel}, \citenamefont {Safronova},
  \citenamefont {Safronova},\ and\ \citenamefont {Porsev}}]{KulFimZip15}%
  \BibitemOpen
  \bibfield  {author} {\bibinfo {author} {\bibfnamefont {A.~P.}\ \bibnamefont
  {Kulosa}}, \bibinfo {author} {\bibfnamefont {D.}~\bibnamefont {Fim}},
  \bibinfo {author} {\bibfnamefont {K.~H.}\ \bibnamefont {Zipfel}}, \bibinfo
  {author} {\bibfnamefont {S.}~\bibnamefont {R\"uhmann}}, \bibinfo {author}
  {\bibfnamefont {S.}~\bibnamefont {Sauer}}, \bibinfo {author} {\bibfnamefont
  {N.}~\bibnamefont {Jha}}, \bibinfo {author} {\bibfnamefont {K.}~\bibnamefont
  {Gibble}}, \bibinfo {author} {\bibfnamefont {W.}~\bibnamefont {Ertmer}},
  \bibinfo {author} {\bibfnamefont {E.~M.}\ \bibnamefont {Rasel}}, \bibinfo
  {author} {\bibfnamefont {M.~S.}\ \bibnamefont {Safronova}}, \bibinfo {author}
  {\bibfnamefont {U.~I.}\ \bibnamefont {Safronova}}, \ and\ \bibinfo {author}
  {\bibfnamefont {S.~G.}\ \bibnamefont {Porsev}},\ }\href {\doibase
  10.1103/PhysRevLett.115.240801} {\bibfield  {journal} {\bibinfo  {journal}
  {Phys. Rev. Lett.}\ }\textbf {\bibinfo {volume} {115}},\ \bibinfo {pages}
  {240801} (\bibinfo {year} {2015})}\BibitemShut {NoStop}%
\bibitem [{\citenamefont {Golovizin}\ \emph {et~al.}(2019)\citenamefont
  {Golovizin}, \citenamefont {Fedorova}, \citenamefont {Tregubov},
  \citenamefont {Sukachev}, \citenamefont {Khabarova}, \citenamefont
  {Sorokin},\ and\ \citenamefont {Kolachevsky}}]{GolFedTre19}%
  \BibitemOpen
  \bibfield  {author} {\bibinfo {author} {\bibfnamefont {A.}~\bibnamefont
  {Golovizin}}, \bibinfo {author} {\bibfnamefont {E.}~\bibnamefont {Fedorova}},
  \bibinfo {author} {\bibfnamefont {D.}~\bibnamefont {Tregubov}}, \bibinfo
  {author} {\bibfnamefont {D.}~\bibnamefont {Sukachev}}, \bibinfo {author}
  {\bibfnamefont {K.}~\bibnamefont {Khabarova}}, \bibinfo {author}
  {\bibfnamefont {V.}~\bibnamefont {Sorokin}}, \ and\ \bibinfo {author}
  {\bibfnamefont {N.}~\bibnamefont {Kolachevsky}},\ }\href {\doibase
  10.1038/s41467-019-09706-9} {\bibfield  {journal} {\bibinfo  {journal} {Nat.
  Comm.}\ }\textbf {\bibinfo {volume} {10}},\ \bibinfo {pages} {1724} (\bibinfo
  {year} {2019})}\BibitemShut {NoStop}%
\bibitem [{\citenamefont {Yamaguchi}\ \emph {et~al.}(2019)\citenamefont
  {Yamaguchi}, \citenamefont {Safronova}, \citenamefont {Gibble},\ and\
  \citenamefont {Katori}}]{YamSafGib19}%
  \BibitemOpen
  \bibfield  {author} {\bibinfo {author} {\bibfnamefont {A.}~\bibnamefont
  {Yamaguchi}}, \bibinfo {author} {\bibfnamefont {M.~S.}\ \bibnamefont
  {Safronova}}, \bibinfo {author} {\bibfnamefont {K.}~\bibnamefont {Gibble}}, \
  and\ \bibinfo {author} {\bibfnamefont {H.}~\bibnamefont {Katori}},\ }\href
  {\doibase 10.1103/PhysRevLett.123.113201} {\bibfield  {journal} {\bibinfo
  {journal} {Phys. Rev. Lett.}\ }\textbf {\bibinfo {volume} {123}},\ \bibinfo
  {pages} {113201} (\bibinfo {year} {2019})}\BibitemShut {NoStop}%
\bibitem [{\citenamefont {Koller}\ \emph {et~al.}(2017)\citenamefont {Koller},
  \citenamefont {Grotti}, \citenamefont {Vogt}, \citenamefont {Al-Masoudi},
  \citenamefont {D\"orscher}, \citenamefont {H\"afner}, \citenamefont {Sterr},\
  and\ \citenamefont {Lisdat}}]{KolGroVog17}%
  \BibitemOpen
  \bibfield  {author} {\bibinfo {author} {\bibfnamefont {S.~B.}\ \bibnamefont
  {Koller}}, \bibinfo {author} {\bibfnamefont {J.}~\bibnamefont {Grotti}},
  \bibinfo {author} {\bibfnamefont {S.}~\bibnamefont {Vogt}}, \bibinfo {author}
  {\bibfnamefont {A.}~\bibnamefont {Al-Masoudi}}, \bibinfo {author}
  {\bibfnamefont {S.}~\bibnamefont {D\"orscher}}, \bibinfo {author}
  {\bibfnamefont {S.}~\bibnamefont {H\"afner}}, \bibinfo {author}
  {\bibfnamefont {U.}~\bibnamefont {Sterr}}, \ and\ \bibinfo {author}
  {\bibfnamefont {C.}~\bibnamefont {Lisdat}},\ }\href {\doibase
  10.1103/PhysRevLett.118.073601} {\bibfield  {journal} {\bibinfo  {journal}
  {Phys. Rev. Lett.}\ }\textbf {\bibinfo {volume} {118}},\ \bibinfo {pages}
  {073601} (\bibinfo {year} {2017})}\BibitemShut {NoStop}%
\bibitem [{\citenamefont {Grotti}\ \emph {et~al.}(2018)\citenamefont {Grotti},
  \citenamefont {Koller}, \citenamefont {Vogt}, \citenamefont {H\"{a}fner},
  \citenamefont {Sterr}, \citenamefont {Lisdat}, \citenamefont {Denker},
  \citenamefont {Voigt}, \citenamefont {Timmen}, \citenamefont {Rolland},
  \citenamefont {Baynes}, \citenamefont {Margolis}, \citenamefont {Zampaolo},
  \citenamefont {Thoumany}, \citenamefont {Pizzocaro}, \citenamefont {Rauf},
  \citenamefont {Bregolin}, \citenamefont {Tampellini}, \citenamefont
  {Barbieri}, \citenamefont {Zucco}, \citenamefont {Costanzo}, \citenamefont
  {Clivati}, \citenamefont {Levi},\ and\ \citenamefont
  {Calonico}}]{GroKolVog18}%
  \BibitemOpen
  \bibfield  {author} {\bibinfo {author} {\bibfnamefont {J.}~\bibnamefont
  {Grotti}}, \bibinfo {author} {\bibfnamefont {S.}~\bibnamefont {Koller}},
  \bibinfo {author} {\bibfnamefont {S.}~\bibnamefont {Vogt}}, \bibinfo {author}
  {\bibfnamefont {S.}~\bibnamefont {H\"{a}fner}}, \bibinfo {author}
  {\bibfnamefont {U.}~\bibnamefont {Sterr}}, \bibinfo {author} {\bibfnamefont
  {C.}~\bibnamefont {Lisdat}}, \bibinfo {author} {\bibfnamefont
  {H.}~\bibnamefont {Denker}}, \bibinfo {author} {\bibfnamefont
  {C.}~\bibnamefont {Voigt}}, \bibinfo {author} {\bibfnamefont
  {L.}~\bibnamefont {Timmen}}, \bibinfo {author} {\bibfnamefont
  {A.}~\bibnamefont {Rolland}}, \bibinfo {author} {\bibfnamefont {F.~N.}\
  \bibnamefont {Baynes}}, \bibinfo {author} {\bibfnamefont {H.~S.}\
  \bibnamefont {Margolis}}, \bibinfo {author} {\bibfnamefont {M.}~\bibnamefont
  {Zampaolo}}, \bibinfo {author} {\bibfnamefont {P.}~\bibnamefont {Thoumany}},
  \bibinfo {author} {\bibfnamefont {M.}~\bibnamefont {Pizzocaro}}, \bibinfo
  {author} {\bibfnamefont {B.}~\bibnamefont {Rauf}}, \bibinfo {author}
  {\bibfnamefont {F.}~\bibnamefont {Bregolin}}, \bibinfo {author}
  {\bibfnamefont {A.}~\bibnamefont {Tampellini}}, \bibinfo {author}
  {\bibfnamefont {P.}~\bibnamefont {Barbieri}}, \bibinfo {author}
  {\bibfnamefont {M.}~\bibnamefont {Zucco}}, \bibinfo {author} {\bibfnamefont
  {G.~A.}\ \bibnamefont {Costanzo}}, \bibinfo {author} {\bibfnamefont
  {C.}~\bibnamefont {Clivati}}, \bibinfo {author} {\bibfnamefont
  {F.}~\bibnamefont {Levi}}, \ and\ \bibinfo {author} {\bibfnamefont
  {D.}~\bibnamefont {Calonico}},\ }\href {\doibase 10.1038/s41567-017-0042-3}
  {\bibfield  {journal} {\bibinfo  {journal} {Nat. Phys.}\ }\textbf {\bibinfo
  {volume} {14}},\ \bibinfo {pages} {437} (\bibinfo {year} {2018})}\BibitemShut
  {NoStop}%
\bibitem [{\citenamefont {Ushijima}\ \emph {et~al.}(2015)\citenamefont
  {Ushijima}, \citenamefont {Takamoto}, \citenamefont {Das}, \citenamefont
  {Ohkubo},\ and\ \citenamefont {Katori}}]{UshTakDas15}%
  \BibitemOpen
  \bibfield  {author} {\bibinfo {author} {\bibfnamefont {I.}~\bibnamefont
  {Ushijima}}, \bibinfo {author} {\bibfnamefont {M.}~\bibnamefont {Takamoto}},
  \bibinfo {author} {\bibfnamefont {M.}~\bibnamefont {Das}}, \bibinfo {author}
  {\bibfnamefont {T.}~\bibnamefont {Ohkubo}}, \ and\ \bibinfo {author}
  {\bibfnamefont {H.}~\bibnamefont {Katori}},\ }\href {\doibase
  10.1038/nphoton.2015.5} {\bibfield  {journal} {\bibinfo  {journal} {Nat.
  Phot.}\ }\textbf {\bibinfo {volume} {9}},\ \bibinfo {pages} {185} (\bibinfo
  {year} {2015})}\BibitemShut {NoStop}%
\bibitem [{\citenamefont {McGrew}\ \emph {et~al.}(2018)\citenamefont {McGrew},
  \citenamefont {Zhang}, \citenamefont {Fasano}, \citenamefont {Sch\"{a}ffer},
  \citenamefont {Beloy}, \citenamefont {Nicolodi}, \citenamefont {Brown},
  \citenamefont {Hinkley}, \citenamefont {Milani}, \citenamefont {Schioppo},
  \citenamefont {Yoon},\ and\ \citenamefont {Ludlow}}]{McGZhaFas18}%
  \BibitemOpen
  \bibfield  {author} {\bibinfo {author} {\bibfnamefont {W.~F.}\ \bibnamefont
  {McGrew}}, \bibinfo {author} {\bibfnamefont {X.}~\bibnamefont {Zhang}},
  \bibinfo {author} {\bibfnamefont {R.~J.}\ \bibnamefont {Fasano}}, \bibinfo
  {author} {\bibfnamefont {S.~A.}\ \bibnamefont {Sch\"{a}ffer}}, \bibinfo
  {author} {\bibfnamefont {K.}~\bibnamefont {Beloy}}, \bibinfo {author}
  {\bibfnamefont {D.}~\bibnamefont {Nicolodi}}, \bibinfo {author}
  {\bibfnamefont {R.~C.}\ \bibnamefont {Brown}}, \bibinfo {author}
  {\bibfnamefont {N.}~\bibnamefont {Hinkley}}, \bibinfo {author} {\bibfnamefont
  {G.}~\bibnamefont {Milani}}, \bibinfo {author} {\bibfnamefont
  {M.}~\bibnamefont {Schioppo}}, \bibinfo {author} {\bibfnamefont {T.~H.}\
  \bibnamefont {Yoon}}, \ and\ \bibinfo {author} {\bibfnamefont {A.~D.}\
  \bibnamefont {Ludlow}},\ }\href {\doibase 10.1038/s41586-018-0738-2}
  {\bibfield  {journal} {\bibinfo  {journal} {Nature}\ }\textbf {\bibinfo
  {volume} {564}},\ \bibinfo {pages} {87} (\bibinfo {year} {2018})}\BibitemShut
  {NoStop}%
\bibitem [{\citenamefont {Bothwell}\ \emph {et~al.}(2019)\citenamefont
  {Bothwell}, \citenamefont {Kedar}, \citenamefont {Oelker}, \citenamefont
  {Robinson}, \citenamefont {Bromley}, \citenamefont {Tew}, \citenamefont
  {Ye},\ and\ \citenamefont {Kennedy}}]{BotKedOel19}%
  \BibitemOpen
  \bibfield  {author} {\bibinfo {author} {\bibfnamefont {T.}~\bibnamefont
  {Bothwell}}, \bibinfo {author} {\bibfnamefont {D.}~\bibnamefont {Kedar}},
  \bibinfo {author} {\bibfnamefont {E.}~\bibnamefont {Oelker}}, \bibinfo
  {author} {\bibfnamefont {J.~M.}\ \bibnamefont {Robinson}}, \bibinfo {author}
  {\bibfnamefont {S.~L.}\ \bibnamefont {Bromley}}, \bibinfo {author}
  {\bibfnamefont {W.~L.}\ \bibnamefont {Tew}}, \bibinfo {author} {\bibfnamefont
  {J.}~\bibnamefont {Ye}}, \ and\ \bibinfo {author} {\bibfnamefont {C.~J.}\
  \bibnamefont {Kennedy}},\ }\href {\doibase 10.1088/1681-7575/ab4089}
  {\bibfield  {journal} {\bibinfo  {journal} {Metrologia}\ }\textbf {\bibinfo
  {volume} {56}},\ \bibinfo {pages} {065004} (\bibinfo {year}
  {2019})}\BibitemShut {NoStop}%
\bibitem [{\citenamefont {Schioppo}\ \emph {et~al.}(2017)\citenamefont
  {Schioppo}, \citenamefont {Brown}, \citenamefont {McGrew}, \citenamefont
  {Hinkley}, \citenamefont {Fasano}, \citenamefont {Beloy}, \citenamefont
  {Yoon}, \citenamefont {Milani}, \citenamefont {Nicolodi}, \citenamefont
  {Sherman}, \citenamefont {Phillips}, \citenamefont {Oates},\ and\
  \citenamefont {Ludlow}}]{SchBroMcG17}%
  \BibitemOpen
  \bibfield  {author} {\bibinfo {author} {\bibfnamefont {M.}~\bibnamefont
  {Schioppo}}, \bibinfo {author} {\bibfnamefont {R.~C.}\ \bibnamefont {Brown}},
  \bibinfo {author} {\bibfnamefont {W.~F.}\ \bibnamefont {McGrew}}, \bibinfo
  {author} {\bibfnamefont {N.}~\bibnamefont {Hinkley}}, \bibinfo {author}
  {\bibfnamefont {R.~J.}\ \bibnamefont {Fasano}}, \bibinfo {author}
  {\bibfnamefont {K.}~\bibnamefont {Beloy}}, \bibinfo {author} {\bibfnamefont
  {T.~H.}\ \bibnamefont {Yoon}}, \bibinfo {author} {\bibfnamefont
  {G.}~\bibnamefont {Milani}}, \bibinfo {author} {\bibfnamefont
  {D.}~\bibnamefont {Nicolodi}}, \bibinfo {author} {\bibfnamefont {J.~A.}\
  \bibnamefont {Sherman}}, \bibinfo {author} {\bibfnamefont {N.~B.}\
  \bibnamefont {Phillips}}, \bibinfo {author} {\bibfnamefont {C.~W.}\
  \bibnamefont {Oates}}, \ and\ \bibinfo {author} {\bibfnamefont {A.~D.}\
  \bibnamefont {Ludlow}},\ }\href {\doibase 10.1038/nphoton.2016.231}
  {\bibfield  {journal} {\bibinfo  {journal} {Nat. Phot.}\ }\textbf {\bibinfo
  {volume} {11}},\ \bibinfo {pages} {48} (\bibinfo {year} {2017})}\BibitemShut
  {NoStop}%
\bibitem [{\citenamefont {Campbell}\ \emph {et~al.}(2017)\citenamefont
  {Campbell}, \citenamefont {Hutson}, \citenamefont {Marti}, \citenamefont
  {Goban}, \citenamefont {Darkwah~Oppong}, \citenamefont {McNally},
  \citenamefont {Sonderhouse}, \citenamefont {Robinson}, \citenamefont {Zhang},
  \citenamefont {Bloom},\ and\ \citenamefont {Ye}}]{CamHutMar17}%
  \BibitemOpen
  \bibfield  {author} {\bibinfo {author} {\bibfnamefont {S.~L.}\ \bibnamefont
  {Campbell}}, \bibinfo {author} {\bibfnamefont {R.~B.}\ \bibnamefont
  {Hutson}}, \bibinfo {author} {\bibfnamefont {G.~E.}\ \bibnamefont {Marti}},
  \bibinfo {author} {\bibfnamefont {A.}~\bibnamefont {Goban}}, \bibinfo
  {author} {\bibfnamefont {N.}~\bibnamefont {Darkwah~Oppong}}, \bibinfo
  {author} {\bibfnamefont {R.~L.}\ \bibnamefont {McNally}}, \bibinfo {author}
  {\bibfnamefont {L.}~\bibnamefont {Sonderhouse}}, \bibinfo {author}
  {\bibfnamefont {J.~M.}\ \bibnamefont {Robinson}}, \bibinfo {author}
  {\bibfnamefont {W.}~\bibnamefont {Zhang}}, \bibinfo {author} {\bibfnamefont
  {B.~J.}\ \bibnamefont {Bloom}}, \ and\ \bibinfo {author} {\bibfnamefont
  {J.}~\bibnamefont {Ye}},\ }\href {\doibase 10.1126/science.aam5538}
  {\bibfield  {journal} {\bibinfo  {journal} {Science}\ }\textbf {\bibinfo
  {volume} {358}},\ \bibinfo {pages} {90} (\bibinfo {year} {2017})}\BibitemShut
  {NoStop}%
\bibitem [{\citenamefont {Beloy}\ \emph {et~al.}(2014)\citenamefont {Beloy},
  \citenamefont {Hinkley}, \citenamefont {Phillips}, \citenamefont {Sherman},
  \citenamefont {Schioppo}, \citenamefont {Lehman}, \citenamefont {Feldman},
  \citenamefont {Hanssen}, \citenamefont {Oates},\ and\ \citenamefont
  {Ludlow}}]{BelHinPhi14}%
  \BibitemOpen
  \bibfield  {author} {\bibinfo {author} {\bibfnamefont {K.}~\bibnamefont
  {Beloy}}, \bibinfo {author} {\bibfnamefont {N.}~\bibnamefont {Hinkley}},
  \bibinfo {author} {\bibfnamefont {N.~B.}\ \bibnamefont {Phillips}}, \bibinfo
  {author} {\bibfnamefont {J.~A.}\ \bibnamefont {Sherman}}, \bibinfo {author}
  {\bibfnamefont {M.}~\bibnamefont {Schioppo}}, \bibinfo {author}
  {\bibfnamefont {J.}~\bibnamefont {Lehman}}, \bibinfo {author} {\bibfnamefont
  {A.}~\bibnamefont {Feldman}}, \bibinfo {author} {\bibfnamefont {L.~M.}\
  \bibnamefont {Hanssen}}, \bibinfo {author} {\bibfnamefont {C.~W.}\
  \bibnamefont {Oates}}, \ and\ \bibinfo {author} {\bibfnamefont {A.~D.}\
  \bibnamefont {Ludlow}},\ }\href {\doibase 10.1103/PhysRevLett.113.260801}
  {\bibfield  {journal} {\bibinfo  {journal} {Phys. Rev. Lett.}\ }\textbf
  {\bibinfo {volume} {113}},\ \bibinfo {pages} {260801} (\bibinfo {year}
  {2014})}\BibitemShut {NoStop}%
\bibitem [{\citenamefont {Beloy}\ \emph {et~al.}(2018)\citenamefont {Beloy},
  \citenamefont {Zhang}, \citenamefont {McGrew}, \citenamefont {Hinkley},
  \citenamefont {Yoon}, \citenamefont {Nicolodi}, \citenamefont {Fasano},
  \citenamefont {Sch\"affer}, \citenamefont {Brown},\ and\ \citenamefont
  {Ludlow}}]{BelZhaMcG18}%
  \BibitemOpen
  \bibfield  {author} {\bibinfo {author} {\bibfnamefont {K.}~\bibnamefont
  {Beloy}}, \bibinfo {author} {\bibfnamefont {X.}~\bibnamefont {Zhang}},
  \bibinfo {author} {\bibfnamefont {W.~F.}\ \bibnamefont {McGrew}}, \bibinfo
  {author} {\bibfnamefont {N.}~\bibnamefont {Hinkley}}, \bibinfo {author}
  {\bibfnamefont {T.~H.}\ \bibnamefont {Yoon}}, \bibinfo {author}
  {\bibfnamefont {D.}~\bibnamefont {Nicolodi}}, \bibinfo {author}
  {\bibfnamefont {R.~J.}\ \bibnamefont {Fasano}}, \bibinfo {author}
  {\bibfnamefont {S.~A.}\ \bibnamefont {Sch\"affer}}, \bibinfo {author}
  {\bibfnamefont {R.~C.}\ \bibnamefont {Brown}}, \ and\ \bibinfo {author}
  {\bibfnamefont {A.~D.}\ \bibnamefont {Ludlow}},\ }\href {\doibase
  10.1103/PhysRevLett.120.183201} {\bibfield  {journal} {\bibinfo  {journal}
  {Phys. Rev. Lett.}\ }\textbf {\bibinfo {volume} {120}},\ \bibinfo {pages}
  {183201} (\bibinfo {year} {2018})}\BibitemShut {NoStop}%
\bibitem [{\citenamefont {Taichenachev}\ \emph {et~al.}(2008)\citenamefont
  {Taichenachev}, \citenamefont {Yudin}, \citenamefont {Ovsiannikov},
  \citenamefont {Pal'chikov},\ and\ \citenamefont {Oates}}]{TaiYudOvs08}%
  \BibitemOpen
  \bibfield  {author} {\bibinfo {author} {\bibfnamefont {A.~V.}\ \bibnamefont
  {Taichenachev}}, \bibinfo {author} {\bibfnamefont {V.~I.}\ \bibnamefont
  {Yudin}}, \bibinfo {author} {\bibfnamefont {V.~D.}\ \bibnamefont
  {Ovsiannikov}}, \bibinfo {author} {\bibfnamefont {V.~G.}\ \bibnamefont
  {Pal'chikov}}, \ and\ \bibinfo {author} {\bibfnamefont {C.~W.}\ \bibnamefont
  {Oates}},\ }\href {\doibase 10.1103/PhysRevLett.101.193601} {\bibfield
  {journal} {\bibinfo  {journal} {Phys. Rev. Lett.}\ }\textbf {\bibinfo
  {volume} {101}},\ \bibinfo {pages} {193601} (\bibinfo {year}
  {2008})}\BibitemShut {NoStop}%
\bibitem [{\citenamefont {Brusch}\ \emph {et~al.}(2006)\citenamefont {Brusch},
  \citenamefont {Le~Targat}, \citenamefont {Baillard}, \citenamefont
  {Fouch\'e},\ and\ \citenamefont {Lemonde}}]{BruLeTBai06}%
  \BibitemOpen
  \bibfield  {author} {\bibinfo {author} {\bibfnamefont {A.}~\bibnamefont
  {Brusch}}, \bibinfo {author} {\bibfnamefont {R.}~\bibnamefont {Le~Targat}},
  \bibinfo {author} {\bibfnamefont {X.}~\bibnamefont {Baillard}}, \bibinfo
  {author} {\bibfnamefont {M.}~\bibnamefont {Fouch\'e}}, \ and\ \bibinfo
  {author} {\bibfnamefont {P.}~\bibnamefont {Lemonde}},\ }\href {\doibase
  10.1103/PhysRevLett.96.103003} {\bibfield  {journal} {\bibinfo  {journal}
  {Phys. Rev. Lett.}\ }\textbf {\bibinfo {volume} {96}},\ \bibinfo {pages}
  {103003} (\bibinfo {year} {2006})}\BibitemShut {NoStop}%
\bibitem [{\citenamefont {Barber}\ \emph {et~al.}(2008)\citenamefont {Barber},
  \citenamefont {Stalnaker}, \citenamefont {Lemke}, \citenamefont {Poli},
  \citenamefont {Oates}, \citenamefont {Fortier}, \citenamefont {Diddams},
  \citenamefont {Hollberg}, \citenamefont {Hoyt}, \citenamefont
  {Taichenachev},\ and\ \citenamefont {Yudin}}]{BarStaLem08}%
  \BibitemOpen
  \bibfield  {author} {\bibinfo {author} {\bibfnamefont {Z.~W.}\ \bibnamefont
  {Barber}}, \bibinfo {author} {\bibfnamefont {J.~E.}\ \bibnamefont
  {Stalnaker}}, \bibinfo {author} {\bibfnamefont {N.~D.}\ \bibnamefont
  {Lemke}}, \bibinfo {author} {\bibfnamefont {N.}~\bibnamefont {Poli}},
  \bibinfo {author} {\bibfnamefont {C.~W.}\ \bibnamefont {Oates}}, \bibinfo
  {author} {\bibfnamefont {T.~M.}\ \bibnamefont {Fortier}}, \bibinfo {author}
  {\bibfnamefont {S.~A.}\ \bibnamefont {Diddams}}, \bibinfo {author}
  {\bibfnamefont {L.}~\bibnamefont {Hollberg}}, \bibinfo {author}
  {\bibfnamefont {C.~W.}\ \bibnamefont {Hoyt}}, \bibinfo {author}
  {\bibfnamefont {A.~V.}\ \bibnamefont {Taichenachev}}, \ and\ \bibinfo
  {author} {\bibfnamefont {V.~I.}\ \bibnamefont {Yudin}},\ }\href {\doibase
  10.1103/PhysRevLett.100.103002} {\bibfield  {journal} {\bibinfo  {journal}
  {Phys. Rev. Lett.}\ }\textbf {\bibinfo {volume} {100}},\ \bibinfo {pages}
  {103002} (\bibinfo {year} {2008})}\BibitemShut {NoStop}%
\bibitem [{\citenamefont {Katori}\ \emph {et~al.}(2009)\citenamefont {Katori},
  \citenamefont {Hashiguchi}, \citenamefont {Il'inova},\ and\ \citenamefont
  {Ovsiannikov}}]{KatHasIli09}%
  \BibitemOpen
  \bibfield  {author} {\bibinfo {author} {\bibfnamefont {H.}~\bibnamefont
  {Katori}}, \bibinfo {author} {\bibfnamefont {K.}~\bibnamefont {Hashiguchi}},
  \bibinfo {author} {\bibfnamefont {E.~Y.}\ \bibnamefont {Il'inova}}, \ and\
  \bibinfo {author} {\bibfnamefont {V.~D.}\ \bibnamefont {Ovsiannikov}},\
  }\href {\doibase 10.1103/PhysRevLett.103.153004} {\bibfield  {journal}
  {\bibinfo  {journal} {Phys. Rev. Lett.}\ }\textbf {\bibinfo {volume} {103}},\
  \bibinfo {pages} {153004} (\bibinfo {year} {2009})}\BibitemShut {NoStop}%
\bibitem [{\citenamefont {Westergaard}\ \emph {et~al.}(2011)\citenamefont
  {Westergaard}, \citenamefont {Lodewyck}, \citenamefont {Lorini},
  \citenamefont {Lecallier}, \citenamefont {Burt}, \citenamefont {Zawada},
  \citenamefont {Millo},\ and\ \citenamefont {Lemonde}}]{WesLodLor11}%
  \BibitemOpen
  \bibfield  {author} {\bibinfo {author} {\bibfnamefont {P.~G.}\ \bibnamefont
  {Westergaard}}, \bibinfo {author} {\bibfnamefont {J.}~\bibnamefont
  {Lodewyck}}, \bibinfo {author} {\bibfnamefont {L.}~\bibnamefont {Lorini}},
  \bibinfo {author} {\bibfnamefont {A.}~\bibnamefont {Lecallier}}, \bibinfo
  {author} {\bibfnamefont {E.~A.}\ \bibnamefont {Burt}}, \bibinfo {author}
  {\bibfnamefont {M.}~\bibnamefont {Zawada}}, \bibinfo {author} {\bibfnamefont
  {J.}~\bibnamefont {Millo}}, \ and\ \bibinfo {author} {\bibfnamefont
  {P.}~\bibnamefont {Lemonde}},\ }\href {\doibase
  10.1103/PhysRevLett.106.210801} {\bibfield  {journal} {\bibinfo  {journal}
  {Phys. Rev. Lett.}\ }\textbf {\bibinfo {volume} {106}},\ \bibinfo {pages}
  {210801} (\bibinfo {year} {2011})}\BibitemShut {NoStop}%
\bibitem [{\citenamefont {Ovsiannikov}\ \emph {et~al.}(2013)\citenamefont
  {Ovsiannikov}, \citenamefont {Pal'chikov}, \citenamefont {Taichenachev},
  \citenamefont {Yudin},\ and\ \citenamefont {Katori}}]{OvsPalTai13}%
  \BibitemOpen
  \bibfield  {author} {\bibinfo {author} {\bibfnamefont {V.~D.}\ \bibnamefont
  {Ovsiannikov}}, \bibinfo {author} {\bibfnamefont {V.~G.}\ \bibnamefont
  {Pal'chikov}}, \bibinfo {author} {\bibfnamefont {A.~V.}\ \bibnamefont
  {Taichenachev}}, \bibinfo {author} {\bibfnamefont {V.~I.}\ \bibnamefont
  {Yudin}}, \ and\ \bibinfo {author} {\bibfnamefont {H.}~\bibnamefont
  {Katori}},\ }\href {\doibase 10.1103/PhysRevA.88.013405} {\bibfield
  {journal} {\bibinfo  {journal} {Phys. Rev. A}\ }\textbf {\bibinfo {volume}
  {88}},\ \bibinfo {pages} {013405} (\bibinfo {year} {2013})}\BibitemShut
  {NoStop}%
\bibitem [{\citenamefont {Katori}\ \emph {et~al.}(2015)\citenamefont {Katori},
  \citenamefont {Ovsiannikov}, \citenamefont {Marmo},\ and\ \citenamefont
  {Palchikov}}]{KatOvsMar15}%
  \BibitemOpen
  \bibfield  {author} {\bibinfo {author} {\bibfnamefont {H.}~\bibnamefont
  {Katori}}, \bibinfo {author} {\bibfnamefont {V.~D.}\ \bibnamefont
  {Ovsiannikov}}, \bibinfo {author} {\bibfnamefont {S.~I.}\ \bibnamefont
  {Marmo}}, \ and\ \bibinfo {author} {\bibfnamefont {V.~G.}\ \bibnamefont
  {Palchikov}},\ }\href {\doibase 10.1103/PhysRevA.91.052503} {\bibfield
  {journal} {\bibinfo  {journal} {Phys. Rev. A}\ }\textbf {\bibinfo {volume}
  {91}},\ \bibinfo {pages} {052503} (\bibinfo {year} {2015})}\BibitemShut
  {NoStop}%
\bibitem [{\citenamefont {Ovsiannikov}\ \emph {et~al.}(2016)\citenamefont
  {Ovsiannikov}, \citenamefont {Marmo}, \citenamefont {Palchikov},\ and\
  \citenamefont {Katori}}]{OvsMarPal16}%
  \BibitemOpen
  \bibfield  {author} {\bibinfo {author} {\bibfnamefont {V.~D.}\ \bibnamefont
  {Ovsiannikov}}, \bibinfo {author} {\bibfnamefont {S.~I.}\ \bibnamefont
  {Marmo}}, \bibinfo {author} {\bibfnamefont {V.~G.}\ \bibnamefont
  {Palchikov}}, \ and\ \bibinfo {author} {\bibfnamefont {H.}~\bibnamefont
  {Katori}},\ }\href {\doibase 10.1103/PhysRevA.93.043420} {\bibfield
  {journal} {\bibinfo  {journal} {Phys. Rev. A}\ }\textbf {\bibinfo {volume}
  {93}},\ \bibinfo {pages} {043420} (\bibinfo {year} {2016})}\BibitemShut
  {NoStop}%
\bibitem [{\citenamefont {Brown}\ \emph {et~al.}(2017)\citenamefont {Brown},
  \citenamefont {Phillips}, \citenamefont {Beloy}, \citenamefont {McGrew},
  \citenamefont {Schioppo}, \citenamefont {Fasano}, \citenamefont {Milani},
  \citenamefont {Zhang}, \citenamefont {Hinkley}, \citenamefont {Leopardi},
  \citenamefont {Yoon}, \citenamefont {Nicolodi}, \citenamefont {Fortier},\
  and\ \citenamefont {Ludlow}}]{BroPhiBel17}%
  \BibitemOpen
  \bibfield  {author} {\bibinfo {author} {\bibfnamefont {R.~C.}\ \bibnamefont
  {Brown}}, \bibinfo {author} {\bibfnamefont {N.~B.}\ \bibnamefont {Phillips}},
  \bibinfo {author} {\bibfnamefont {K.}~\bibnamefont {Beloy}}, \bibinfo
  {author} {\bibfnamefont {W.~F.}\ \bibnamefont {McGrew}}, \bibinfo {author}
  {\bibfnamefont {M.}~\bibnamefont {Schioppo}}, \bibinfo {author}
  {\bibfnamefont {R.~J.}\ \bibnamefont {Fasano}}, \bibinfo {author}
  {\bibfnamefont {G.}~\bibnamefont {Milani}}, \bibinfo {author} {\bibfnamefont
  {X.}~\bibnamefont {Zhang}}, \bibinfo {author} {\bibfnamefont
  {N.}~\bibnamefont {Hinkley}}, \bibinfo {author} {\bibfnamefont
  {H.}~\bibnamefont {Leopardi}}, \bibinfo {author} {\bibfnamefont {T.~H.}\
  \bibnamefont {Yoon}}, \bibinfo {author} {\bibfnamefont {D.}~\bibnamefont
  {Nicolodi}}, \bibinfo {author} {\bibfnamefont {T.~M.}\ \bibnamefont
  {Fortier}}, \ and\ \bibinfo {author} {\bibfnamefont {A.~D.}\ \bibnamefont
  {Ludlow}},\ }\href {\doibase 10.1103/PhysRevLett.119.253001} {\bibfield
  {journal} {\bibinfo  {journal} {Phys. Rev. Lett.}\ }\textbf {\bibinfo
  {volume} {119}},\ \bibinfo {pages} {253001} (\bibinfo {year}
  {2017})}\BibitemShut {NoStop}%
\bibitem [{\citenamefont {Ushijima}\ \emph {et~al.}(2018)\citenamefont
  {Ushijima}, \citenamefont {Takamoto},\ and\ \citenamefont
  {Katori}}]{UshTakKat18}%
  \BibitemOpen
  \bibfield  {author} {\bibinfo {author} {\bibfnamefont {I.}~\bibnamefont
  {Ushijima}}, \bibinfo {author} {\bibfnamefont {M.}~\bibnamefont {Takamoto}},
  \ and\ \bibinfo {author} {\bibfnamefont {H.}~\bibnamefont {Katori}},\ }\href
  {\doibase 10.1103/PhysRevLett.121.263202} {\bibfield  {journal} {\bibinfo
  {journal} {Phys. Rev. Lett.}\ }\textbf {\bibinfo {volume} {121}},\ \bibinfo
  {pages} {263202} (\bibinfo {year} {2018})}\BibitemShut {NoStop}%
\bibitem [{\citenamefont {Nemitz}\ \emph {et~al.}(2019)\citenamefont {Nemitz},
  \citenamefont {J\o{}rgensen}, \citenamefont {Yanagimoto}, \citenamefont
  {Bregolin},\ and\ \citenamefont {Katori}}]{NemJorYan19}%
  \BibitemOpen
  \bibfield  {author} {\bibinfo {author} {\bibfnamefont {N.}~\bibnamefont
  {Nemitz}}, \bibinfo {author} {\bibfnamefont {A.~A.}\ \bibnamefont
  {J\o{}rgensen}}, \bibinfo {author} {\bibfnamefont {R.}~\bibnamefont
  {Yanagimoto}}, \bibinfo {author} {\bibfnamefont {F.}~\bibnamefont
  {Bregolin}}, \ and\ \bibinfo {author} {\bibfnamefont {H.}~\bibnamefont
  {Katori}},\ }\href {\doibase 10.1103/PhysRevA.99.033424} {\bibfield
  {journal} {\bibinfo  {journal} {Phys. Rev. A}\ }\textbf {\bibinfo {volume}
  {99}},\ \bibinfo {pages} {033424} (\bibinfo {year} {2019})}\BibitemShut
  {NoStop}%
\bibitem [{\citenamefont {Bloch}\ and\ \citenamefont
  {Zoller}(2006)}]{BloZol06}%
  \BibitemOpen
  \bibfield  {author} {\bibinfo {author} {\bibfnamefont {I.}~\bibnamefont
  {Bloch}}\ and\ \bibinfo {author} {\bibfnamefont {P.}~\bibnamefont {Zoller}},\
  }\href {\doibase 10.1088/1367-2630/8/8/E02} {\bibfield  {journal} {\bibinfo
  {journal} {New J. Phys.}\ }\textbf {\bibinfo {volume} {8}} (\bibinfo {year}
  {2006}),\ 10.1088/1367-2630/8/8/E02}\BibitemShut {NoStop}%
\bibitem [{\citenamefont {Lemonde}\ and\ \citenamefont
  {Wolf}(2005)}]{LemWol05}%
  \BibitemOpen
  \bibfield  {author} {\bibinfo {author} {\bibfnamefont {P.}~\bibnamefont
  {Lemonde}}\ and\ \bibinfo {author} {\bibfnamefont {P.}~\bibnamefont {Wolf}},\
  }\href {\doibase 10.1103/PhysRevA.72.033409} {\bibfield  {journal} {\bibinfo
  {journal} {Phys. Rev. A}\ }\textbf {\bibinfo {volume} {72}},\ \bibinfo
  {pages} {033409} (\bibinfo {year} {2005})}\BibitemShut {NoStop}%
\bibitem [{myf()}]{myfootnote}%
  \BibitemOpen
  \href@noop {} {}\bibinfo {note} {{It is worth noting that, consistent with
  previous studies, a Born-Oppenheimer-type approximation has already been
  invoked (albeit implicitly) at the outset of the problem. Specifically,
  external and internal degrees of freedom of the atom take the respective
  roles of the nuclear and electronic degrees of freedom of the molecular
  problem. The potential $U(\rho,z)$, which depends on the internal state of
  the atom via the $E1$ polarizability, is analogous to potential energy curves
  (or surfaces) of the molecular problem. Reference~\cite{Bel10} examines the
  breakdown of this approximation in optical lattice clocks.}}\BibitemShut
  {Stop}%
\bibitem [{\citenamefont {Berry}\ and\ \citenamefont
  {de~Almeida}(1973)}]{BerOzo73}%
  \BibitemOpen
  \bibfield  {author} {\bibinfo {author} {\bibfnamefont {M.~V.}\ \bibnamefont
  {Berry}}\ and\ \bibinfo {author} {\bibfnamefont {A.~M.~O.}\ \bibnamefont
  {de~Almeida}},\ }\href {\doibase 10.1088/0305-4470/6/10/005} {\bibfield
  {journal} {\bibinfo  {journal} {J. Phys. A}\ }\textbf {\bibinfo {volume}
  {6}},\ \bibinfo {pages} {1451} (\bibinfo {year} {1973})}\BibitemShut
  {NoStop}%
\bibitem [{\citenamefont {Blatt}\ \emph {et~al.}(2009)\citenamefont {Blatt},
  \citenamefont {Thomsen}, \citenamefont {Campbell}, \citenamefont {Ludlow},
  \citenamefont {Swallows}, \citenamefont {Martin}, \citenamefont {Boyd},\ and\
  \citenamefont {Ye}}]{BlaThoCam09}%
  \BibitemOpen
  \bibfield  {author} {\bibinfo {author} {\bibfnamefont {S.}~\bibnamefont
  {Blatt}}, \bibinfo {author} {\bibfnamefont {J.~W.}\ \bibnamefont {Thomsen}},
  \bibinfo {author} {\bibfnamefont {G.~K.}\ \bibnamefont {Campbell}}, \bibinfo
  {author} {\bibfnamefont {A.~D.}\ \bibnamefont {Ludlow}}, \bibinfo {author}
  {\bibfnamefont {M.~D.}\ \bibnamefont {Swallows}}, \bibinfo {author}
  {\bibfnamefont {M.~J.}\ \bibnamefont {Martin}}, \bibinfo {author}
  {\bibfnamefont {M.~M.}\ \bibnamefont {Boyd}}, \ and\ \bibinfo {author}
  {\bibfnamefont {J.}~\bibnamefont {Ye}},\ }\href {\doibase
  10.1103/PhysRevA.80.052703} {\bibfield  {journal} {\bibinfo  {journal} {Phys.
  Rev. A}\ }\textbf {\bibinfo {volume} {80}},\ \bibinfo {pages} {052703}
  (\bibinfo {year} {2009})}\BibitemShut {NoStop}%
\bibitem [{\citenamefont {Porsev}\ \emph {et~al.}(2018)\citenamefont {Porsev},
  \citenamefont {Safronova}, \citenamefont {Safronova},\ and\ \citenamefont
  {Kozlov}}]{PorSafSaf18}%
  \BibitemOpen
  \bibfield  {author} {\bibinfo {author} {\bibfnamefont {S.~G.}\ \bibnamefont
  {Porsev}}, \bibinfo {author} {\bibfnamefont {M.~S.}\ \bibnamefont
  {Safronova}}, \bibinfo {author} {\bibfnamefont {U.~I.}\ \bibnamefont
  {Safronova}}, \ and\ \bibinfo {author} {\bibfnamefont {M.~G.}\ \bibnamefont
  {Kozlov}},\ }\href {\doibase 10.1103/PhysRevLett.120.063204} {\bibfield
  {journal} {\bibinfo  {journal} {Phys. Rev. Lett.}\ }\textbf {\bibinfo
  {volume} {120}},\ \bibinfo {pages} {063204} (\bibinfo {year}
  {2018})}\BibitemShut {NoStop}%
\bibitem [{\citenamefont {Wu}\ \emph {et~al.}(2019)\citenamefont {Wu},
  \citenamefont {Tang}, \citenamefont {Shi},\ and\ \citenamefont
  {Tang}}]{WuTanShi19}%
  \BibitemOpen
  \bibfield  {author} {\bibinfo {author} {\bibfnamefont {F.-F.}\ \bibnamefont
  {Wu}}, \bibinfo {author} {\bibfnamefont {Y.-B.}\ \bibnamefont {Tang}},
  \bibinfo {author} {\bibfnamefont {T.-Y.}\ \bibnamefont {Shi}}, \ and\
  \bibinfo {author} {\bibfnamefont {L.-Y.}\ \bibnamefont {Tang}},\ }\href
  {\doibase 10.1103/PhysRevA.100.042514} {\bibfield  {journal} {\bibinfo
  {journal} {Phys. Rev. A}\ }\textbf {\bibinfo {volume} {100}},\ \bibinfo
  {pages} {042514} (\bibinfo {year} {2019})}\BibitemShut {NoStop}%
\bibitem [{\citenamefont {Beloy}()}]{Bel19unpublished}%
  \BibitemOpen
  \bibfield  {author} {\bibinfo {author} {\bibfnamefont {K.}~\bibnamefont
  {Beloy}},\ }\href@noop {} {\ }\bibinfo {note} {{unpublished}}\BibitemShut
  {NoStop}%
\bibitem [{rep()}]{reparamnote}%
  \BibitemOpen
  \href@noop {} {}\bibinfo {note} {{References~\cite{BroPhiBel17,McGZhaFas18}
  consider the error associated with reparametrization of the lattice light
  shift. In the supplemental material of Ref.~\cite{BroPhiBel17}, it is
  referred to as ``lattice shift model error.'' In Table 1 of
  Ref.~\cite{McGZhaFas18}, it is referred to as ``lattice light (model).''
  }}\BibitemShut {NoStop}%
\bibitem [{\citenamefont {Akatsuka}\ \emph {et~al.}(2008)\citenamefont
  {Akatsuka}, \citenamefont {Takamoto},\ and\ \citenamefont
  {Katori}}]{AkaTakKat08}%
  \BibitemOpen
  \bibfield  {author} {\bibinfo {author} {\bibfnamefont {T.}~\bibnamefont
  {Akatsuka}}, \bibinfo {author} {\bibfnamefont {M.}~\bibnamefont {Takamoto}},
  \ and\ \bibinfo {author} {\bibfnamefont {H.}~\bibnamefont {Katori}},\ }\href
  {\doibase 10.1038/nphys1108} {\bibfield  {journal} {\bibinfo  {journal} {Nat.
  Phys.}\ }\textbf {\bibinfo {volume} {4}},\ \bibinfo {pages} {954} (\bibinfo
  {year} {2008})}\BibitemShut {NoStop}%
\bibitem [{\citenamefont {Blanch}(1972)}]{Bra72}%
  \BibitemOpen
  \bibfield  {author} {\bibinfo {author} {\bibfnamefont {G.}~\bibnamefont
  {Blanch}},\ }\href@noop {} {\emph {\bibinfo {title} {\emph{in} Handbook of
  Mathematical Functions with Formulas, Graphs, and Mathematical Tables}}},\
  \bibinfo {edition} {10th}\ ed.,\ edited by\ \bibinfo {editor} {\bibfnamefont
  {M.}~\bibnamefont {Abramowitz}}\ and\ \bibinfo {editor} {\bibfnamefont
  {I.~A.}\ \bibnamefont {Stegun}}\ (\bibinfo  {publisher} {U.S. Government
  Printing Office},\ \bibinfo {year} {1972})\ p.\ \bibinfo {pages}
  {721}\BibitemShut {NoStop}%
\bibitem [{\citenamefont {Beloy}(2010)}]{Bel10}%
  \BibitemOpen
  \bibfield  {author} {\bibinfo {author} {\bibfnamefont {K.}~\bibnamefont
  {Beloy}},\ }\href {\doibase 10.1103/PhysRevA.82.031402} {\bibfield  {journal}
  {\bibinfo  {journal} {Phys. Rev. A}\ }\textbf {\bibinfo {volume} {82}},\
  \bibinfo {pages} {031402} (\bibinfo {year} {2010})}\BibitemShut {NoStop}%
\end{thebibliography}

%

\end{document}